\documentclass[iop,apj]{emulateapj}
\usepackage{amsmath}

\newcommand\msol{{\cal M_{\odot}}}
\newcommand\teff{{T_{\rm eff}}}

\newcommand\pab{$\langle P_{ab}\rangle$}

\newcommand\delv{$\Delta V_{TO}^{HB}$} 
\newcommand\amlt{\alpha_{\rm MLT}}
\newcommand\lta{\mathrel{\hbox{\raise 0.6 ex \hbox{$<$}\kern
                   -1.8 ex\lower .5 ex\hbox{$\sim$}}}}
\newcommand\gta{\mathrel{\hbox{\raise 0.6 ex \hbox{$>$}\kern
                   -1.7 ex\lower .5 ex\hbox{$\sim$}}}}

\begin{document}

\epsscale{1.1}

\title{CONSTRAINTS ON THE DISTANCE MODULI, HELIUM AND METAL ABUNDANCES, AND
AGES OF GLOBULAR CLUSTERS FROM THEIR RR LYRAE AND NON-VARIABLE HORIZONTAL-BRANCH
STARS.~III.~M\,55 AND NGC\,6362}

\shortauthors{VandenBerg \& Denissenkov}
\shorttitle{HB Stars in M\,55 and NGC\,6362}

\author{Don A.~VandenBerg and P.~A.~Denissenkov} 
\affil{Department of Physics \& Astronomy, University of Victoria,
 P.O.~Box 1700 STN CSC, Victoria, B.C.\ \ V8W~2Y2, Canada;\hfil\break
 vandenbe@uvic.ca, pavelden@uvic.ca}


\submitted{Submitted to The Astrophysical Journal}

\begin{abstract}

M\,55 (NGC\,6809) and NGC 6362 are among the few globular clusters for which
masses and radii have been derived to high precision for member binary stars.
They also contain RR Lyrae variables which, together with their non-variable 
horizontal-branch (HB) populations, provide tight constraints on the cluster
reddenings and distance moduli through fits of stellar models to their
pulsational and evolutionary properties.  Reliable $(m-M)_V$ estimates yield
$M_V$ and $M_{\rm bol}$ values of comparable accuracy for binary stars because
the $V$-band bolometric corrections applicable to them have no more than a weak
dependence on effective temperature ($\teff$) and [Fe/H].  Chemical abundances
derived from the binary mass--$M_V$ relations are independent of determinations
based on their spectra.  The temperatures of the binaries, which are calculated
directly from their luminosities and the measured radii, completely rule out the
low $\teff$\ scale that has been determined for metal-deficient stars in some
recent spectroscopic and interferometric studies.  If [$\alpha$/Fe] $= 0.4$ and
[O/Fe] $= 0.5 \pm 0.1$, we find that M\,55 has $(m-M)_V = 13.95 \pm 0.05$,
[Fe/H] $= -1.85 \pm 0.1$, and an age of $12.9 \pm 0.8$\ Gyr, whereas NGC\,6362
has $(m-M)_V = 14.56 \pm 0.05$, [Fe/H] $= -0.90 \pm 0.1$, and an age of $12.4
\pm 0.8$ Gyr.  The HB of NGC\,6362 shows clear evidence for multiple stellar
populations.  Constraints from the RR Lyrae standard candle and from local
subdwarfs (with {\it Gaia} DR2 parallaxes) are briefly discussed.

\end{abstract}


\section{Introduction}
\label{sec:intro}

More than 60 years have passed since photometry derived from photographic plates
taken at the 200-inch telescope on Mt.~Palomar revealed for the first time the
turnoffs of globular clusters (GCs) --- specifically, those of M\,92
(\citealt{abs53}) and M\,3 (\citealt{san53}).  During the following decade,
color-magnitude diagrams (CMDs) extending down to the main sequence (MS) were
obtained for several other GCs, including M\,13 (\citealt{bhj59}), M\,5
(\citealt{arp62}), and 47 Tuc (\citealt{tif63}).  Interestingly, the basic
properties of these clusters that were derived for them (in particular, their
distances and metallicities) are closer to present-day determinations than one
might have expected. 

For instance, most of the pioneering studies mentioned in the previous paragraph
argued in support of the apparent distance moduli, $(m-M)_V$, that were obtained
if RR Lyrae variables were assumed to have $M_V \approx 0.6$.  This was
supported, in part, by the determination of $M_V = 0.65\pm 0.1$ for RR Lyr
itself from the application of the moving-group method (see \citealt{es59}).
However, \citet{san58} suspected early on that cluster-to-cluster variations
in the mean periods of the RR Lyrae could be explained if the luminosity of the
horizontal branch (HB) increased with decreasing metallicity.  Subsequently,
\citet{sw60} suggested that it would be reasonable to place the HB at $M_V =
0.4$ in M\,92, at $M_V = 0.6$ in M\,3 (to be consistent with RR Lyr), and at
fainter absolute magnitudes in more metal-rich systems.  As it turns out,
distance moduli inferred from recent models for the HB phase imply $\langle
M_V\rangle = 0.35$ and 0.58 for M\,92 and M\,3, respectively (see Table 1 in
\citealt[hereafter Paper I]{vdc16}), which agree with the results from the
Sandage \& Wallerstein paper to within several hundredths of a magnitude.

Indeed, the main advance during the intervening years has been to reduce the
uncertainties associated with the derived $(m-M)_V$ values from $\sim\pm
0.2$--0.3 mag (e.g., \citealt{bhj59}, \citealt{tif63}) to $\sim\pm 0.1$ mag.
In well observed clusters such as M\,92, M\,5, and 47\,Tuc, the $1\,\sigma$
uncertainties appear to be somewhat less than this; see Table~\ref{tab:t1},
which lists many of the apparent distance moduli that have been derived for
these three clusters since the turn of the century.  As indicated in the second
column, some studies used local subdwarf (SBDWF) or subgiant branch (SGB) stars,
RR Lyrae (RRL), or white dwarfs (WD) as standard candles.  Others constrained
the cluster distances using luminosity functions (LFs), red-giant (RG) clump or
tip stars, eclipsing binary members, RR Lyrae period-luminosity (PL) relations,
theoretical results that give the RR Lyrae pulsation period as a function of its
mass, luminosity, effective temperature, and metallicity (${\cal M}L\teff Z$),
or detailed comparisons between synthetic and observed HB populations (``HB
fits"). Some of the earliest estimates of $(m-M)_V$ --- e.g., 14.62 for M\,92 by
\citet{sw66}, 14.39 for M\,5 by \citet{arp62}, and 13.35 for 47\,Tuc by
\citet{tif63} --- clearly agree quite well with the tabulated values.

\begin{table}[t]
\begin{center}
\caption{Determinations of $(m-M)_V$ for M\,92, M\,5, and 47\,Tuc}
\label{tab:t1}
\begin{tabular}{lcc}
\hline
\hline
\noalign{\vskip 1pt}
Reference & Method & $(m-M)_V$ \\ [0.5ex]
\hline
\noalign{\vskip 1pt}
\multicolumn{3}{c}{\bf M\,92} \\ [0.5ex]
\citet{cgc00}      &  SBDWF                   & 14.72 \\
\citet{van00}      &  ZAHB                    & 14.70 \\
\citet{vrm02}      &   SGB                    & 14.62 \\
\citet{dpb05}      & RRL (PL)                 & 14.67 \\
\citet{scv06}      & RRL (PL)                 & 14.73 \\
\citet{pcs07}      &   LFs                    & 14.66 \\
\citet{apm09}      &  SBDWF                   & 14.71 \\
\citet{bmf11}      &   RRL                    & 14.78 \\
\citet{vbl13}      &  ZAHB                    & 14.72 \\
\citet{vdc16}      & RRL (${\cal M}L\teff Z$) & 14.74 \\ 
\citet{cmo17}      &  SBDWF                   & 14.89 \\ [0.5ex]
Average($\sigma$)  &                          & $14.72(0.07)$ \\ [0.5ex]
\multicolumn{3}{c}{\bf M\,5} \\ [0.5ex]
\citet{cgc00}      &   SBDWF                  & 14.57 \\
\citet{van00}      &   ZAHB                   & 14.48 \\
\citet{dmc04}      & RRL (${\cal M}L\teff Z$) & 14.41 \\
\citet{lsv05}      &   SBDWF                  & 14.56 \\
\citet{scv06}      & RRL (PL)                 & 14.46 \\
\citet{apm09}      &   SBDWF                  & 14.40 \\
\citet{cdr11}      &  RRL (PL)                & 14.53 \\
\citet{vbl13}      &   ZAHB                   & 14.38 \\
\citet{vbn14}      &  SBDWF                   & 14.40 \\
\citet{alb16}      &  RRL (PL)                & 14.49 \\ [0.5ex]
Average($\sigma$)  &                          & $14.47(0.07)$ \\ [0.5ex]
\multicolumn{3}{c}{\bf 47\,Tuc} \\ [0.5ex]
\citet{cgc00}      &   SBDWF                  & 13.55 \\
\citet{van00}      &   ZAHB                   & 13.37 \\
\citet{fmo00}      &   RG tip                 & 13.44 \\
\citet{zro01}      &    WD                    & 13.27 \\
\citet{psv02}      &   SBDWF                  & 13.37 \\
\citet{gsa02}      &   SBDWF                  & 13.33 \\
\citet{sg02}       &  RG clump                & 13.34 \\
\citet{bs09}       &   SBDWF                  & 13.38 \\
\citet{tkr10}      &  binary                  & 13.35 \\
\citet{scp16}      &   HB fits                & 13.40 \\
\citet{dvk17}      &   HB fits                & 13.27 \\
\citet{bvb17}      &  binary                  & 13.30 \\ [0.5ex]
Average($\sigma$)  &                          & $13.37(0.08)$ \\ [0.5ex]
\hline
\end{tabular}
\end{center}
\end{table}

Similarly, the overall metallicities that were derived for GCs in the 1960s
seem quite reasonable from today's perspective, even though little was known
about the detailed metals mixtures at the time.  For example, based on
measurements of the ultraviolet excesses in cluster stars, \citet{arp62}
concluded that the metals-to-hydrogen ($M$/H) ratios in M\,92 and M\,5 differed
from the solar value by, in turn, factors of $\lta 1/200$ and $\sim 1/17$,
whereas spectra of individual giants in 47 Tuc implied a factor $\gta 1/4$
(\citealt{ft60}).  In the usual logarithmic notation, the corresponding [$M$/H]
values are $-2.3$, $-1.2$, and $-0.6$, which are $\lta 0.15$ dex higher than the
[Fe/H] values that have been obtained in modern spectroscopic studies for M\,92,
M\,5, and 47\,Tuc, respectively (e.g., \citealt{cg97}; \citealt{ki03};
\citealt[hereafter CBG09]{cbg09a}).  (Differences of a similar amount, but in the opposite sense,
are implied by current [$M$/H] values that take into account enhanced abundances
of the $\alpha$-elements by 0.3--0.4 dex.) 

\begin{figure}[t]
\plotone{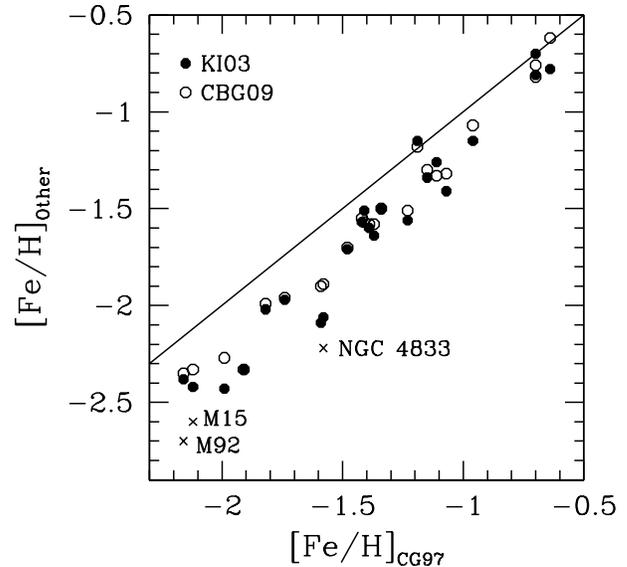}
\caption{Comparison of the [Fe/H] values for 23 globular clusters derived by
\citet[CG97]{cg97} with those given by \citet[KI03, {\it filled circles}]{ki03},
CBG09 ({\it open circles}), and recent studies ({\it crosses}) of
M\,15 (\citealt{sks11}), M\,92 (\citealt{rs11}), and NGC\,4833 (\citealt{rt15}).
Note that the vertical offsets of some of the points from the diagonal ``line
of equality" are $\gta 0.6$ dex.} 
\label{fig:feh}
\end{figure}

However, despite the vast amount of spectroscopic work that has been carried out
over the years (see, e.g., the compilations of published results by
\citealt{pvi05}, \citealt{rcg14}), the uncertainties associated with the
{\it absolute} abundances of many metals remain large.  This is illustrated in
Figure~\ref{fig:feh}, which shows that the [Fe/H] values derived by
\citet[hereafter CG97]{cg97} tend to be $\gta 0.2$ dex larger than those found
by \citet[hereafter KI03]{ki03}, except at the metal-rich end.  High-resolution
spectra were used in both investigations, but Kraft \& Ivans anchored their
determinations to [Fe/H] values based on Fe\,II lines to minimize the effects
of possible departures from local thermodynamic equilibrium (LTE), whereas
neutral iron lines provided the basis of the CG97 metallicity scale.  A
recalibration of the latter by CBG09, using even higher
resolution spectra, improved $gf$ values, and a somewhat cooler $\teff$ scale,
resulted in [Fe/H] determinations that agreed reasonably well with those by
KI03, though differences at the level of 0.1--0.2 dex persisted for many
clusters (note the vertical separations between the pairs of open and filled
circles in Fig.~\ref{fig:feh}).

Unfortunately, even more recent spectroscopic studies have tended to ``muddy the
water".  The latest results for M\,15 (\citealt{sks11}), M\,92 (\citealt{rs11}),
and NGC\,4833 (\citealt{rt15}), which have been plotted as crosses in
Fig.~\ref{fig:feh}, are $\sim 0.3$ dex lower than the CBG09 determinations, and
$\sim 0.6$ dex less than the values reported by CG97.  As discussed by Roederer
\& Sneden, the factor of two reduction in the iron abundances relative to the
findings that some of the same authors had previously obtained (e.g.,
\citealt{spk00}) appears to be due to differences in (i) the Fe\,I lines that
were selected for analysis, (ii) the adopted $gf$ values, (iii) the treatment
of the Rayleigh scattering component of the blue continuous opacity, and (iv)
the 1D model atmospheres that were employed.

Even when extremely high quality spectra are available, as in the case of nearby
halo stars with well determined distances, the [Fe/H] values derived from their
spectra can vary by $> 0.3$ dex.  This is exemplified by recent
work on the turnoff (TO) star HD\,84937, for which \citet{ala16} obtained
[Fe/H] $= -1.96$ from detailed 3D, non-LTE radiative transfer calculations, as
compared with the value of [Fe/H] $= -2.32$ that was found by \citet{sck16}
using 1D model atmospheres.  (The latter argue that there are no substantial
departures from LTE in this star; for additional discussion on this point and
possible concerns with 3D model atmospheres at low metallicities, see
\citealt{spg17}.)  

\begin{table}[t]
\begin{center}
\caption{Properties of HD\,84937}
\label{tab:t2}
\begin{tabular}{lccc}
\hline
\hline
\noalign{\vskip 1pt}
Reference & $\log\,g$ & $T_{\rm eff}$ & [Fe/H] \\ [0.5ex]
\hline
\citet{gcc96}      & 4.06 & 6344 & $-2.04$ \\
\citet{jeb05}      & 4.04 & 6310 & $-1.96$ \\
\citet{gsz06}      & 4.00 & 6346 & $-2.16$ \\
\citet{cps07}      & 4.01 & 6228 & $-2.17$ \\
\citet{mzg08}      & 4.00 & 6365 & $-2.15$ \\
\citet{crm10}      & 3.93 & 6408 & $-2.11$ \\
\citet{blc12}      & 4.13 & 6408 & $-2.16$ \\
\citet{ral13}      & 4.15 & 6377 & $-2.02$ \\
\citet{vbn14}      & 4.05 & 6408 & $-2.08$ \\
\citet{szm15}      & 4.09 & 6350 & $-2.12$ \\
\citet{ala16}      & 4.06 & 6356 & $-1.96$ \\
\citet{sck16}      & 4.00 & 6300 & $-2.32$ \\ [0.5ex]
\hline
\end{tabular}
\end{center}
\end{table}

Most studies of HD\,84937 over the past two decades have found intermediate
[Fe/H] values --- as shown in Table~\ref{tab:t2}, which lists only a
representative fraction ($\lta 50$\%) of the investigations that considered this
star during the past two decades.  Although the metallicity that was reported in
any one of the tabulated papers has some dependence on the adopted temperature,
the $\teff$\ value seems to be less important than other ingredients of
chemical abundance determinations (model atmospheres, atomic physics, etc.).
For instance, some of the studies referenced in Table~\ref{tab:t2} adopted
nearly the same $\teff$\ values and yet their [Fe/H] determinations differ by
$> 0.3$\ dex (see the entries for \citealt{jeb05}, \citealt{sck16}), while
others found nearly the same metallicities, despite assuming very different
temperatures (e.g., \citealt{cps07}, \citealt{blc12}).  Judging from the
discordant results that were published $\sim 2$~years ago (\citealt{ala16},
\citealt{sck16}), it may be some time before we can claim that the
{\it absolute} metallicity of HD\,84937 is known to better than 0.1 dex.
(Considering just the tabulated findings, the average [Fe/H] value is $-2.10$,
with a $1\,\sigma$ standard deviation from the mean of 0.10 dex.)

At the present time, the relatively high temperatures that are found from the
application of the infrared flux method (IRFM; see, e.g., \citealt[hereafter
CRMBA]{crm10}; \citealt{mcr10}) seem to be favored (also from the theoretical
perspective; see the overlays of isochrones onto the CMD locations of nearby
subdwarfs by \citealt{vcs10}), but this issue is not yet settled.  As shown by
CRMBA, IRFM temperatures agree reasonably well with those derived from the
hydrogen lines by \citet{ber08} and \citet{fna09}, except at $\teff \gta 6100$~K
where they are hotter by $\sim 50$--100~K.  The recent analysis of
interferometric data for the nearby, [Fe/H] $\sim -2.4$ subgiant, HD\,140283,
by \citet{ctb15} may also be a potential problem for the IRFM $\teff$\ scale,
but as discussed in their paper, stellar models would require a very low value
of the mixing-length parameter ($\amlt$) in order to explain their observations.
This would be at odds with the implications of globular cluster CMDs for $\amlt$
(see Paper I) as well as recent calibrations of the mixing-length
parameter based on 3D hydrodynamical model atmospheres (\citealt{mwa15}).
Consequently, it is not clear whether HD\,140283 has anomalous properties or
the temperature derived for it by Creevey et al.~is too low.  In any case, even
if the IRFM $\teff$\ scale is trustworthy, the typical uncertainties of
$\sim\pm 70$~K for a given star still imply a range of {\it most probable}
values that spans nearly 150~K.

Uncertainties associated with stellar temperatures and {\it absolute} [Fe/H]
values (as well as [O/H], [Mg/H], etc.; see, e.g., \citealt{fna09}; 
\citealt{ral13}; \citealt[\citealt{bcs17}]{blc12}; \citealt[and references
therein]{zmy16}), which appear to range from $\sim\pm 0.1$--0.25 dex (especially
at the lowest metallicities), obviously limit our ability to test and to improve
stellar models.  (Much higher precisions are quoted in most abundance
determinations, but the stated uncertainties will not have taken into account
the errors associated with such things as the assumed atmospheric temperature
structures, the adopted atomic physics, and the evaluation of non-LTE effects,
which are not easily determined.)  Discrepancies between predicted and observed
CMDs, for instance, can easily be due to problems with the predicted $\teff$
values, given that they are very dependent on the treatment of convection and
the atmospheric boundary condition (see, e.g., \citealt[2014]{vee08}).  However,
it is also possible that errors in the photometry, the adopted color--$\teff$
relations, and/or the assumed cluster properties (reddening, distance, chemical
abundances) are responsible.  Unless the basic properties of the stars are known
to high accuracy, it is very difficult to evaluate, e.g., the reliability of
different color transformations, the extent to which $\amlt$\ varies with mass,
metallicity, or evolutionary state, etc.  
  
Tightening the constraints on the empirical metallicity and $\teff$\ scales
would certainly help to break through the current impasse.  The presence of
detached, eclipsing binaries in GCs that also contain RR Lyrae provides an
avenue for doing just that.  Normally, binaries have been used to obtain an
independent estimate of the age and distance modulus of a given GC on the
assumption of spectroscopically derived abundances and temperatures of the
components that are usually inferred from their colors (see, e.g.,
\citealt{tkr10}, \citealt{ktr13}).  Although distance-independent ages can be
derived from the binary mass--radius and mass--$M_V$ relations, they are not
particularly well constrained because the predicted radii depend on uncertain
model temperatures, and the luminosities of the binary components, from which
their $M_V$ values are calculated, depend on the uncertain $\teff$\ values that
are adopted for them.  This has been demonstrated in the study of the 47\,Tuc
binary, V69, by \citet{bvb17}, who have also shown that the helium and metal
abundances are important variables in such analyses, because lower $Y$ has
similar effects on mass--radius and mass--luminosity relations at a fixed age
as increased [Fe/H] and/or [O/Fe].

Previous papers in the present series have given us considerable confidence that
accurate distances of GCs, especially those that contain RR Lyrae variables, can
be determined from state-of-the-art HB models.  To be more specific, we showed
in Paper I that the observed periods of the RR
Lyrae in M\,3, M\,15, and M\,92 agree very well with those predicted by modern
HB tracks on the assumption of distance moduli that are obtained from fits of
ZAHB models to the non-variable HB populations of each cluster.  The distance
modulus obtained in this way for M\,3, in particular, is in excellent agreement
with that implied by the best available calibration of the RR Lyrae standard
candle at [Fe/H] $\sim -1.5$.  Moreover, the synthetic HB populations that were
generated by \citet[hereafter Paper II]{dvk17} provided very realistic
reproductions of the observed distributions of HB stars in M\,3, M\,13, and
47\,Tuc, if the helium abundance varies by $\delta Y \approx 0.02$, 0.08, and
0.03, respectively.  Thus, HB models are able to place tight limits on He
abundance variations within GCs (as already demonstrated by, e.g.,
\citealt{scp16}), as well as on their distances.

Such results are not particularly dependent on the adopted metallicities.  That
is, assuming a higher or lower [Fe/H] by $\lta 0.2$ dex would entail an
adjustment of the ZAHB-based distance modulus by only $\lta \pm 0.04$ mag, and
simply by making a small, compensating offset of the model $\teff$\ scale (by
$\delta\log\teff \lta 0.004$), one would obtain essentially the same fit to the 
pulsational properties of the RR Lyrae.  (The $\delta\teff$\ estimate in this
example is based on the reasonable assumptions that $\delta M_{\rm bol} \approx
\delta M_V$, and that pulsational periods, which vary directly with luminosity
but inversely with temperature, are nearly four times as dependent on
$\log\teff$ as on $\log(L/L_\odot)$; see \citealt{mcb15}.)

However, once the distance modulus is set, and the consequent TO age is
determined, the isochrone for that age must be able to reproduce the binary
mass--$M_V$ relation.  In general, some iteration of
the assumed abundances and/or the adopted value of $(m-M)_V$ will be necessary
to achieve a consistent interpretation of both the member binaries and the
cluster RR Lyrae (if, indeed, it is possible to do so).  If consistency is
obtained, then the effective temperatures of the binary components that are
calculated directly from the measured radii and the derived luminosities
will provide valuable constraints on the stellar $\teff$ scale at the [Fe/H]
value of the GC under consideration.  

In principle, detached, ecliping binaries in GCs provide a particularly
promising way to determine the temperatures of metal-poor stars over a very
wide range in [Fe/H].  Since the radii of their components can often be measured
to within 1\%, the uncertainties in the $\teff$\ values that are derived from
$L = 4\pi R^2\sigma\teff^4$ are mainly limited by the accuracy of the intrinsic
luminosities.  In order for the temperatures derived in this way to be
competitive with those based on alternative methods, it is necessary to know the
distance modulus of the binary, and hence of the cluster that it resides in, to
better than $\sim\pm 0.08$ mag, as this translates into $\delta\log\teff\approx
\pm 0.01$ if $\delta R/R \sim 0.01$ and the errors in the apparent magnitudes
are $\lta 0.01$--0.02 mag.  Moreover, binary mass-luminosity relations,
which are nearly independent of model atmospheres and synthetic spectra (aside
from the conversion of $M_V$ to $M_{\rm bol}$), provide an important
constraint of the chemical abundances that are derived spectroscopically.

In this investigation, the approach described above is applied to the globular
clusters M\,55 (NGC\,6809) and NGC\,6362, which have iron abundances that
differ by about a factor of 10 (see, e.g., CBG09).  The masses and
radii of the detached eclipsing binary in M\,55 (known as V54) have
been derived to within 2.1\% and 0.95\%, respectively, by \citet{ktd14}.  The
same group (see \citealt{ktd15}) have been able to measure the masses of the
components of two detached eclipsing binaries, V40 and V41, in NGC\,6362 to
better than $0.71$\%, as compared with radius uncertainties of $< 1.3$\%.
Mean magnitudes and colors of the 15 RR Lyrae variables that
have been found in M\,55 are provided by \citet{okt99}, who also measured these
quantities in the nearly three dozen RR Lyrae that reside in NGC\,6362
(\citealt{okt01}).

The next section briefly describes the stellar models and additional theoretical
results that are used in this work.   Sections 3 and 4 contain, in turn, our
analyses of the CMDs, the RR Lyrae populations, and the eclipsing binaries in
M\,55 and in NGC\,6362.  The implications of the distance moduli that we have
determined from HB models for the calibration of the RR Lyrae standard candle
is presented in section 5, where a fit of the M\,55 main sequence to local
subdwarfs is also reported and discussed.  The main conclusions of this
investigation are summarized in section 6.

\section{Theoretical Considerations}
\label{sec:theory}

All of the stellar evolutionary calculations considered in this paper adopt the
solar mix of heavy elements given by \citet{ags09} as the reference mixture.  At
the [Fe/H] values of interest, enhanced $\alpha$-element abundances by $+0.4$
dex have generally been assumed, which is consistent with the mean values of
[Mg/Fe] and [Si/Fe] that have been derived in the spectroscopic study of 19 GCs,
including M\,55, by \citet{cbg09b}.  Although the latter find that oxygen is
less enhanced, with $\langle$[O/Fe]$\rangle$ closer to $+0.2$ than to $+0.4$, 
star-to-star variations in the abundance of this element are typically $\gta
0.4$ dex; e.g., for the clusters considered by Carretta et al.~(see their Table
10), the mean {\it rms} variation of $\langle$[O/Fe]$\rangle$\ is 0.18 dex.
(Such variations are apparent in the ubiquitous O--Na anticorrelations
that are widely considered to be one of the defining properties of GCs.)  Since
oxygen will be converted to nitrogen as the result of CNO-cycling at high
temperatures, the highest O abundances, which are roughly consistent with
[$\alpha$/Fe] $= +0.4$, are more likely to be representative of the initial
abundance than the mean cluster abundance.\footnote{Isochrones have a
strong dependence on the total C$+$N$+$O abundance, which appears to be nearly
constant within a given GC (see, e.g., \citealt{ssb96}, \citealt{cm05}, \citealt{sci05}),
with little sensitivity to the ratio C:N:O.  For instance, there is no separation
of CN-weak and CN-strong stars in CMDs that are derived from broad-band filters
(see, e.g., \citealt[their Fig.~6]{ccb98}).} In fact, [O/Fe] values in
GCs with $-2.0 \lta$ [Fe/H] $\lta -0.8$ could be as high as $\sim +0.6$ if their
CN-weak populations have the same oxygen abundances as field stars with similar
metallicities (see, e.g., \citealt{fna09}; \citealt[2013]{rmc12};
\citealt{zmy16}).

Isochrones for [$\alpha$/Fe] $= +0.4$ and the values of $Y$ and [Fe/H] that are
relevant to M\,55 and NGC\,6362 have been generated using the interpolation
software and grids of evolutionary tracks provided by \citet{vbf14}.  To examine
how the interpretation of the observations is affected by a change in the
assumed oxygen abundance, we have also made some use of isochrones (to be the
subject of a forthcoming paper, still in preparation) in which the abundances of
all of the $\alpha$ elements, except oxygen, are enhanced by $+0.4$ dex, and
[O/Fe] is treated as a free parameter.  The version of the Victoria code that
has been used to compute all of the above models has been described in
considerable detail by \citet[and references therein]{vbd12}.

Because further development of this computer program is needed in order to
follow the evolution of core helium-burning stars, revision 7624 of the MESA code
(\citealt{pbd11}), with an improved treatment of the mixing at the boundary of
the convective He core (see Paper I), has been used to produce all of the ZAHBs
and HB tracks that are compared with the observed HBs.  (Although the
Victoria and MESA programs produce nearly identical tracks and ZAHB loci when
essentially the same physics is assumed, as shown in Paper I, the version of
the Victoria code that was used to generate the isochrones which are employed in
this study adopted a slightly different treatment of the atmospheric layers than
that assumed in the MESA code.  The main effect of this difference is a small
shift in the predicted $\teff$~scales between the respective model computations,
which is most evident along the RGB where colors have a stronger dependence on
temperature than in the case of bluer stars.  However, this is inconsequential
for the present study as the isochrone colors are usually adjusted by a small
amount (typically $\lta 0.02$ mag) in order to match the observed TO, and
thereby to derive the best estimate of the cluster age for a given value of
$(m-M)_V$.  The Victoria-Regina isochrones generally reproduce the morphologies
of observed CMDs in the vicinity of the TO quite well because, unlike MESA
models, they take into account extra mixing below surface convection zones, when
they exist, to limit the diffusion of the metals from the atmospheric layers
(see \citealt{vbd12}).

\begin{figure}[t]
\plotone{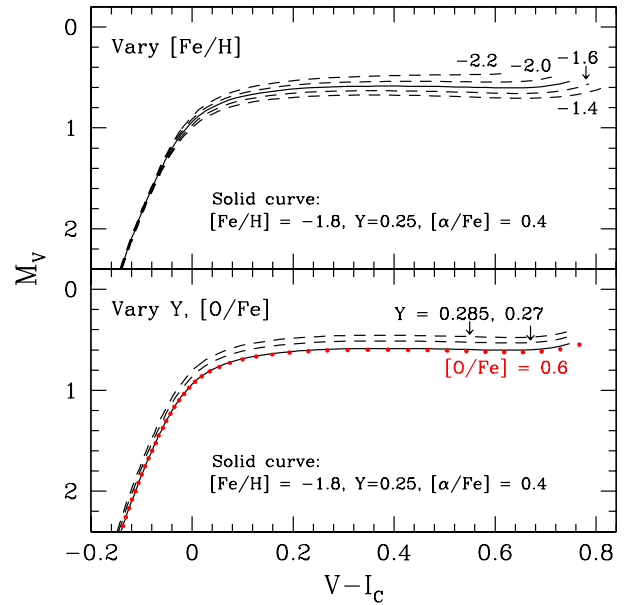}
\caption{Relative to the ZAHB represented by the solid curve, which was computed
for indicated chemical abundances, the effects of varying [Fe/H] are shown in
the top panel, whereas the effects of varying $Y$ and [O/Fe] are illustrated in
the bottom panel.}
\label{fig:hbmet}
\end{figure}

As already mentioned in \S\,\ref{sec:intro}, fits of ZAHB models to the lower
bounds of the distributions of HB stars in GCs yield what appear to be very good
estimates of their respective apparent distance moduli.  In fact, depending on
whether or not the ``knee" of the HB (i.e., the transition from a steeply sloped
blue tail to a much more horizontal morphology at redder colors) is populated,
such fits may also provide tight constraints on the cluster reddening ---
because the location of the blue tail in optical CMDs is nearly independent of
the metallicity.  This is shown in the top panel of Figure~\ref{fig:hbmet}, 
which compares 5 ZAHBs for [Fe/H] values that range from $-2.2$ to $-1.4$, and
by the dotted curve in the bottom panel, which assumes the same chemical
abundances as the solid curve, except for an increased oxygen abundance by 0.2
dex.  Varying [O/Fe] causes a displacement of the red end of a ZAHB to somewhat
redder and fainter colors, but it has no significant effect on the location of
its blue end.  Hence, the distance moduli of metal-poor GCs that are inferred
from fits of ZAHB models to blue HB stars that lie blueward of the instability
strip are independent of the oxygen abundance.  The bottom panel also illustrates
the well known strong sensitivity of the luminosity of the HB to relatively small
changes in the helium abundance.  At the blue end, the increased luminosities
of higher $Y$ ZAHBs have the effect of making the blue tails bluer at a fixed
magnitude, but not by a large amount.  (Plots in the Str\"omgren system
show similar characteristics; see, e.g., \citealt{cfs09}.)

When \citet[hereafter VBLC13]{vbl13} determined the ages of more than 4 dozen
GCs for which \citet{sbc07} had obtained {\it Hubble Space Telescope (HST)}
photometry, they adopted $E(B-V)$\ values from the \citet{sfd98} dust maps and
found that the morphologies of the blue HBs in clusters that possess them were
reproduced exceedingly well by ZAHB models.  Importantly, the distance moduli
implied by the same ZAHB fits were in excellent agreement with those based on
the RR Lyrae standard candle.  Moreover, Paper I has shown that consistent
interpretations of the data are obtained on both {\it HST} and Johnson-Cousins
$BVI_C$ color planes when the same reddenings and distances are adopted.  Since
Schlegel et al.~give $E(B-V) = 0.135$ for M\,55 and $E(B-V) = 0.076$ for
NGC\,6362, we therefore anticipate being able to explain the CMDs and the
properties of the RR Lyrae in these clusters on the assumption of reddenings
that are reasonably close to these values.

As in Papers I and II, the pulsation periods are calculated using 
\begin{align}
\log\,P_{ab} &={}11.347 + 0.860\,\log(L/L_\odot) - 3.43\,\log\teff\nonumber\\
                   &- 0.58\,\log({\cal{M}/\cal{M}_\odot}) + 0.024\,\log\,Z
\end{align}
\noindent and
\begin{align}
\log\,P_c &={}11.167 + 0.822\,\log(L/L_\odot) - 3.40\,\log\teff\nonumber\\ 
           &- 0.56\,\log({\cal{M}/\cal{M}_\odot}) + 0.013\,\log\,Z.
\end{align}
These equations, which have been derived from sophisticated hydrodynamical
models of RR Lyrae variables by \citet{mcb15}, enable one to calculate the
periods of the $ab$-type (fundamental mode) and $c$-type (first overtone)
pulsators if the luminosity, effective temperature, and mass of each variable
is found by interpolating with a grid of HB tracks that was computed for a
metallicity $Z$ (i.e., the total mass-fraction abundance of all elements
heavier than helium).  As reported by Marconi et al., the uncertainties of the
numerical coefficients and the constant terms in equations (1) and (2) are all
quite small.  

We now turn our attention to M\,55, which bears considerable similarity to 
M\,92 insofar as both have predominately blue HBs and relatively few RR Lyrae,
though they have different metallicities by $\delta$[Fe/H] $\gta 0.4$ dex.

\section{M\,55 (NGC\,6809)}
\label{sec:m55}

As is the case for a large fraction of the Galactic GCs, M\,55 has been the
subject of numerous studies over the years (e.g., \citealt{pe84};
\citealt{bbh90}; \citealt{mfr96}; \citealt[hereafter VS07]{vas07};
\citealt{prz10}).  Most estimates of its metallicity have favored [Fe/H]
$\approx -1.8$ (e.g., \citealt{zw84}, KI03, \citealt{khg08}), but support can be
found for both higher and lower values by $\sim\pm 0.15$ dex (see, e.g.,
\citealt{cd88}, \citealt{rhs97}, \citealt{mgp93}, CBG09).  Although $E(B-V) =
0.08$ is listed for M\,55 in the latest edition of the \citet{har96} catalog
(see footnote 1), several studies have found that such a low value is
problematic (see \citealt{svh88}, \citealt{mfr96}, \citealt{ktd14}).
Relatively high values ($\gta 0.12$ mag) are permitted by the line-of-sight
reddenings from dust maps
(\citealt{sfd98}, \citealt{sf11}).  Finally, the apparent distance moduli that
have been derived for M\,55, using a variety of methods, have tended to be in
the range $13.9 \lta (m-M)_V \lta 14.15$ (\citealt{mfr96}, \citealt{pz99},
VS07, VBLC13, \citealt{ktd14}).

For the present work, we decided to investigate the $B-V,\,V$ CMD for this
cluster that was obtained by \citet[hereafter 
KTKZ10]{ktk10},\footnote{http://case.camk.edu.pl/results/Photometry/M55/index.html}
mainly because members of the same group (\citealt{okt99}) had previously
determined intensity-averaged $V$ magnitudes ($\langle V\rangle$) and
$\langle B\rangle -\langle V\rangle$ colors for the RR Lyrae in M\,55. (Note
that colors which are obtained from the difference of intensity-weighted mean
magnitudes appear to be especially good approximations to the colors of
equivalent static stars; see \citealt{bcs95}.)  As discussed by KTKZ10,
their photometry was collected in a number of observing runs between May 1997
and June 2009 and they were calibrated to the standard Johnson-Cousins $BVRI_C$
system using transformations given by \citet{st00}.  However, in contrast with
our findings in the case of M\,3, M\,15, and M\,92  (see Paper I), we were
unable to achieve a fully consistent fit of our ZAHB  models to these $BV$
observations {\it and} the {\it HST} $F606W,\,F814W$ photometry by \citet{sbc07}
for the bluest HB stars in M\,55.  As reported by VBLC13 in their study of 55
of the GCs observed by Sarajedini et al., our ZAHB models generally provide
very satisfactory fits to the morphologies of observed HBs on the assumption of
well supported reddenings and distances.  This led us to wonder if there may be
small zero-point errors in the photometric data provided by KTKZ10.

\begin{figure}[t]
\plotone{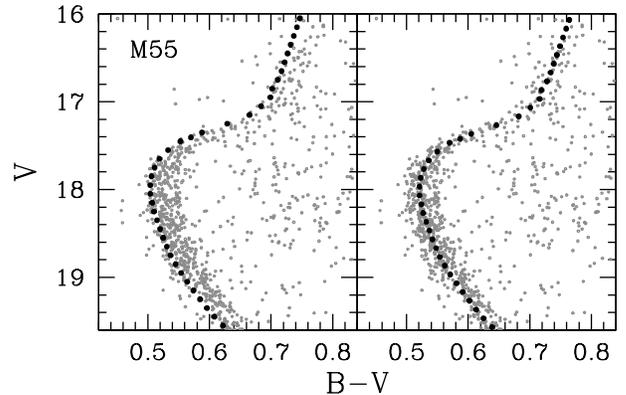}
\caption{{\it Left-hand panel}: comparison of the median fiducial sequence
(black, filled circles) that represents the $B-V,\,V$ CMD of M\,55 by 
\citet{ktk10} with the standard-field photometry (gray points) from P.~Stetson's
archive (see the text for details and the relevant website address).
{\it Right-hand panel}: similar to the left-hand panel, except that the sequence
of black, filled circles has been adjusted by $\delta(B-V) = 0.017$ mag and
$\delta\,V = 0.017$ mag prior to being plotted.}
\label{fig:stdf}
\end{figure}   

One way of checking into this possibility is to compare the CMD procured by
KTKZ10 with the one that can be derived from the publicly available
``Photometric Standard Fields" archive that was developed by \citet{st00} 
and subsequently maintained by
him.\footnote{www.cadc.hia.nrc-cnrc.gc.ca/en/community/STETSON/}
Since this database has been steadily evolving since its inception (see, e.g.,
\citealt{st05}), one can anticipate that the latest calibrations of secondary
cluster standards will differ to some extent from those reported by \citet{st00}.
In fact, as shown in the left-hand panel of Figure~\ref{fig:stdf},
the $B-V,\,V$ diagram for M\,55 from this archive does differ in small, but
significant, ways from the KTKZ10 CMD, which is represented by the small black
filled circles.  (This fiducial sequence consists of the median points in 0.1
mag bins that were determined for the region of the CMD within $\sim\pm 2$\ mag
of the turnoff (TO).  Because there are only about a half-dozen blue HB stars
in Stetson's archive, which show considerable scatter, any zero-point offsets
that might be present are most readily seen by comparing the turnoff photometry
from the two sources.)  In order for the fiducial sequence to reproduce the
archival data, it must be corrected by $\delta(B-V) = 0.017$ mag and $\delta\,V
\approx 0.017$ mag, resulting in the comparison between the two that is
illustrated in the right-hand panel.  (The color adjustment is more reliably
determined and more important than the magnitude offset, which could be larger
or smaller than our estimate by $\sim 0.03$\ mag.) 

\begin{figure}[t]
\plotone{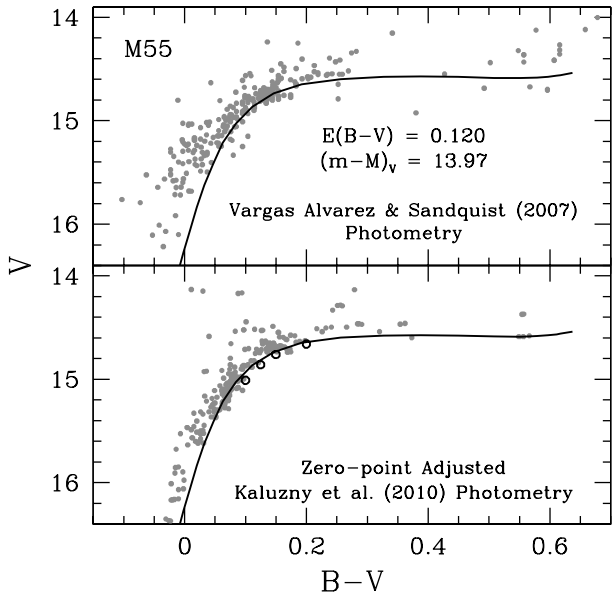}
\caption{{\it Top panel}: Comparison of HB photometry (gray points) for M\,55
from VS07 with a ZAHB for [Fe/H] $= -1.80$, [$\alpha$/Fe] $= +0.4$, and
$Y = 0.25$.  The models were shifted to the observed plane assuming $E(B-V) =
0.12$ and $(m-M)_V = 13.97$.  {\it Bottom panel}: As in the top panel, except
that the zero-point-adjusted photometry from \citet{ktk10} has been plotted
(grey points).  The four open circles indicate the red edge of the distribution
of HB stars in the vicinity of the knee, based on the CMD published by
\citet[see their Fig.~1]{okt99}.  No zero-point corrections were applied to
these data.}
\label{fig:hbvas}
\end{figure}

Independent $BV$ photometry for the HB stars in M\,55 by VS07 provides
important confirmation of these offsets.  This data set contains more HB stars
in the vicinity of the knee, and somewhat larger scatter (see the upper panel
of Figure~\ref{fig:hbvas}) than the KTKZ10 CMD (lower panel).  Nevertheless,
the ZAHB that has been plotted, for [Fe/H] $= -1.80$, [$\alpha$/Fe] $= +0.4$,
and $Y = 0.25$, clearly provides a comparable fit to the observations from the
two sources if the aforementioned adjustments are applied to the KTKZ10 CMD.
(In this, and the next, plot, we have adopted the values of $E(B-V)$ and
$(m-M)_V$ that are favored by our simulations of the cluster HB; see
\S\,\ref{subsec:m55hbsim}.  It turns out that nearly the same reddening,
$E(B-V) = 0.117$, is given by \citet{sf11} from their recalibration of the
\citet{sfd98} dust maps.)

Thus, the offsets that were determined from the TO
observations (see Fig.~\ref{fig:stdf}) are nearly the same as those needed to
obtain consistency with the photometry of the HB by VS07.  In addition, as shown
by the small open circles in the bottom panel, the lower bound to the
distribution of HB stars in the CMD obtained by \citet{okt99} matches its
counterpart in the zero-point-corrected CMD of KTKZ10 quite well.  (Differences
in the calibration of the photometry, as described in the respective papers,
likely explain why there are small offsets between them.)  The open circles were
obtained by producing a magnified version of Fig.~1 in the paper by Olech et
al., drawing a smooth curve through the faintest stars in the vicinity of the
knee of the blue HB, and then determining the $B-V,\,V$ co-ordinates at four
locations on that curve.  We conclude from this exercise that it is unnecessary
to apply any adjustments to the values of $\langle V\rangle$ and
$\langle B\rangle - \langle V\rangle$ that were determined by Olech et al.~for
the cluster RR Lyrae.  (Although not shown, a comparison of the respective
giant-branch loci supports this conclusion.)
 
\begin{figure}[t]
\plotone{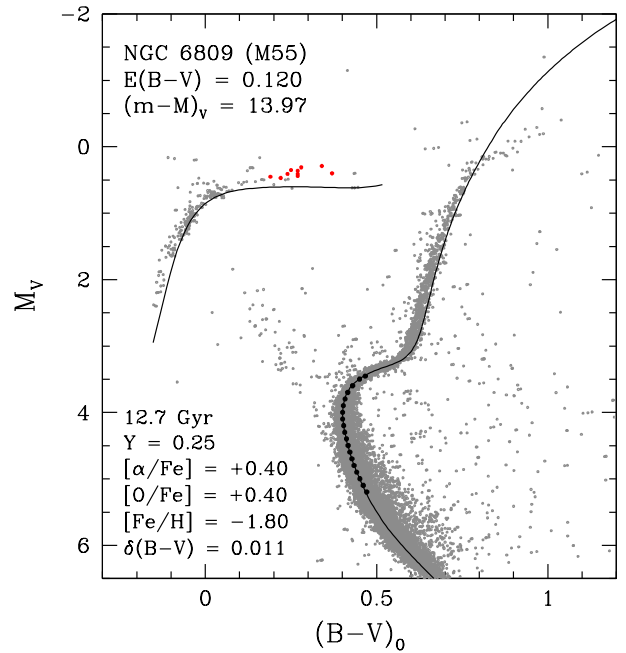}
\caption{Fit of a ZAHB and a 12.7 Gyr isochrone to the HB and turnoff (TO)
observations of M\,55, respectively, by \citet{ktk10}, taking into account small
zero-point adjustments amounting to $\delta(B-V) = 0.017$\ mag and $\delta\,V =
0.017$\ mag; see the text.  (As noted by Kaluzny et al., the images of stars
with $V < 14.0$\ mag were overexposed, which explains the odd morphology of the
upper giant branch.)  The indicated chemical abundances, reddening, and
apparent distance modulus in the $V$ magnitude have been assumed.  The small
filled circles in the vicinity of the TO indicate the median fiducial sequence
through the photometric data.  In order to reproduce the observed TO color, the
isochrone that provides the best simultaneous fit to the TO and the stars just
beginning their subgiant evolution had to be corrected by only 0.011 mag (to the
red).  The small filled circles in red located just above the ZAHB represent
cluster RR Lyrae stars, for which \citet{okt99} have derived intensity-weighted
magnitudes and colors ($\langle V\rangle$ and $\langle B-V\rangle$).}
\label{fig:bvm18}
\end{figure}

Of the 15 RR Lyrae that were observed by \citet{okt99}, two of them (V14 and
V15) are suspected to be members of the Sagittarius dwarf galaxy, while three
others (V9, V10, and V12) have irregular light curves that suggested (to them)
the presence of non-radial oscillations.  (These irregularities could
alternatively be the manifestation of the Blazhko effect, as proposed for
NGC\,6362 variables that show similar anomalies by \citealt{smk17}.)  Of the
remaining 10 RR Lyrae, V2, V4--V8, and V11 were found to be probable cluster
members according to the recent proper motion study by \citet{zkt11}.  It is
not known whether V1, V3, and V13 are members, and even though we have found
irreconcilable differences between the predicted and observed periods of these
stars (see below), there was no reason to exclude them from our sample of M\,55
RR Lyrae at the outset of this work.  In this investigation, we have therefore
considered 10 variables, of which four are fundamental mode pulsators and the
rest are first overtone pulsators.  (To better approximate the colors of
equivalent static stars, the small amplitude-dependent corrections to
$\langle B\rangle - \langle V\rangle$ that were derived by \citet{bcs95} for
$ab$-type RR Lyrae have been taken into account.  According to the latter,
no such adjustments are needed for first overtone pulsators.)  

On the assumption of what should be quite accurate estimates of $E(B-V)$ and
$(m-M)_V$\ (pending our analysis of the periods of the cluster RR Lyrae), we
found that a 12.7 Gyr isochrone for the same chemical abundances ([Fe/H] $=
-1.80$, [$\alpha$/Fe] $= +0.4$, and $Y = 0.25$) does a good job of reproducing
the TO observations.  We checked, and found, that the models yielded
essentially the same interpretation of {\it HST} observations for M\,55
(\citealt{sbc07}), showing that there is rather good consistency between the
two photometric data sets as well as the transformations from the
$(\log\,\teff,\,M_{\rm bol})$-diagram to the observed CMDs. (Because of
its similarity with Fig.~\ref{fig:bvm18}, a plot showing our fit of the same
ZAHB and isochrones to the {\it HST} data has not been included in this paper.)
The small offset between the giant-branch portion of the best-fit isochrone
and the observed RGB could easily be caused by any one or more of the
uncertainties that affect predicted temperatures and colors (some of which are
listed in the last paragraph of this section). 

As explained by VBLC13, only a narrow region of the CMD in the vicinity of the
TO, where the shapes of isochrones are predicted to be nearly independent of,
among other quantities, age and the mixing-length parameter, should be
considered when determining the age corresponding to an adopted distance
modulus.  Using a well defined fiducial sequence to represent the turnoff
observations facilitates the determination of the best-fit isochrone because it
helps to ensure that each isochrone, in turn, is registered to exactly the same
TO color when attempting to identify which one of them also reproduces the
location of the stars just at the beginning of the SGB (those just brighter
and redder than the TO).  Only one isochrone can
provide a simultaneous match of both of these features, for a given value of
$(m-M)_V$, and it is the age of this isochrone that is the best estimate of the
cluster age for the assumed chemical abundances.  Even though a 12.7 Gyr
isochrone clearly provides a very good fit to the median fiducial sequences in
Figs.~\ref{fig:bvm18}, it should be appreciated that an
isochrone for a different age would provide an equally good fit had a larger
or smaller distance modulus by the appropriate amount been assumed.

In order to match the TO photometry, it is generally necessary to correct the
isochrone colors by a small amount that depends to some extent on the cluster
and the filter passbands under consideration (see Papers I and II).  Errors in
the adopted color--$\teff$\ relations, the predicted temperatures (due, e.g.,
to inadequacies in the treatment of convection or the atmospheric boundary
condition), the photometry, and/or the basic properties of a GC (reddening,
distance, chemical composition) can easily explain such problems, but it is very
difficult to determine which source of uncertainty is primarily responsible for
them.  However, it should be kept in mind that ad hoc corrections to the colors
of isochrones have no significant impact on age determinations because the
latter are derived from the magnitude difference between the ZAHB and the TO.

\subsection{Simulations of the M\,55 HB}
\label{subsec:m55hbsim}

As shown in Paper II, simulations of the HBs of observed CMDs, including the RR
Lyrae components, are able to provide tight constraints on not only the cluster
reddening and distance modulus but also the star-to-star variations in the
initial helium abundance.  Using the synthesis code introduced in that paper,
we have generated synthetic HB distributions from newly computed grids of core
He-burning tracks for [Fe/H] $= -1.80$, [$\alpha$/Fe] $= +0.4$, and several
values of $Y$.  These simulations have been compared with the observed
distribution of the HB stars in M\,55, on the assumption that its RR Lyrae have
the colors and magnitudes of equivalent static stars derived from the values
of $\langle B\rangle - \langle V\rangle$ and $\langle V\rangle$ that were 
derived by \citet{okt99}, with the small adjustments given by \citet{bcs95}.  

The best HB fit corresponds to the maximum probability predicted by the
Kolmogorov-Smirnov (K-S) test (see Paper II) when we attempt to reproduce the
observed distributions of the HB stars in both color and
magnitude simultaneously, while varying the parameters in our HB population
synthesis tool.  These parameters include the relative fractions $f_i$ of the
HB stars with different initial masses ${\cal M}_i$ and He abundances $Y_i$
that are used to represent the multiple stellar populations residing in a given
GC, the mean masses that are lost by the RGB stars of these populations
$\Delta {\cal M}_i$, and their standard deviations $\sigma_i$ --- as well as the
GC distance modulus and reddening.  (For this exercise, we have assumed that
the magnitudes of the brightest, unsaturated stars in the KTKZ10 CMD have
$\sigma_{\rm phot} \le 0.003$ mag.)

\begin{figure*}[t]
\plotone{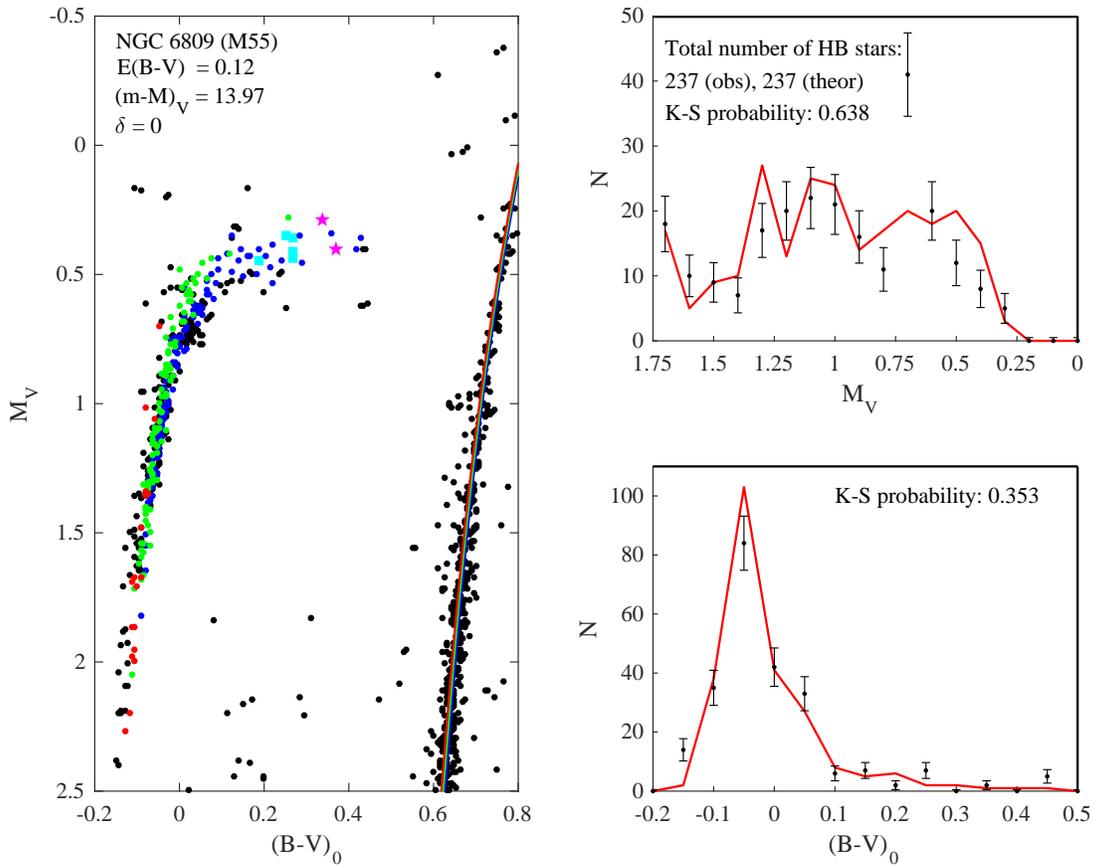}
\caption{{\it Left-hand panel:} overlay of a synthetic HB population of NGC\,6809
 (M\,55) derived from evolutionary tracks for [Fe/H]\,$=-1.80$, [$\alpha$/Fe]\,$
 =0.4$, [O/Fe]\,$=0.4$, and $Y_i = 0.25$, 0.265, and 0.28 (blue, green, and red
 filled circles, respectively) onto the observed CMD (black filled circles).
 RR Lyrae are represented by squares and star symbols.  {\it Right-hand panels:}
 Comparisons of the predicted and observed numbers of HB stars (red loci and
 black dots with error bars, respectively) as a function of $M_V$ and $(B-V)_0$
 color.  The relatively high values of the K-S probability indicate that the
 simulated and observed distributions of stars are essentially equivalent.} 
\label{fig:syn6809}
\end{figure*}

The closest match that we were able to obtain between our synthetic HB
populations and the observed HB in M\,55 is shown in Figure~\ref{fig:syn6809}
for the distance modulus and reddening that are indicated in the left-hand
panel.  Note that these estimates of $(m-M)_V$ and $E(B-V)$ differ by only 0.01
mag from the ZAHB-based determinations discussed in the previous section (see
Figs.~\ref{fig:hbvas} and~\ref{fig:bvm18}), which is well within photometric and
fitting uncertainties.  As in Paper II, the blue, green, and red filled circles
in this panel represent the simulated HB populations with increasing helium
abundances (specifically, $Y_i = 0.25$, 0.265, and 0.28, respectively, in
this case).  The fractions of stars with these helium abundances and the
corresponding values of the synthesis parameters are listed in
Table~\ref{tab:fitm55}.  The squares and star symbols represent the RRc
and RRab variables.

\begin{table}[b]
\begin{center}
\caption{Fitted Parameters of M\,55 from Simulations of its HB}
\label{tab:fitm55}
\begin{tabular}{cccccc}
\hline
\noalign{\vskip 2pt}
\hline
\noalign{\vskip 1pt}
 Population ($i$) & $f_i$ & $Y_i$ & ${\cal M}_i/{\cal M}_\odot$ &
 $\Delta {\cal M}_i/{\cal M}_\odot$ & $\sigma_i/{\cal M}_\odot$ \\ [0.5ex]
\hline
 $1$ & $0.51$ & $0.250$ & $0.787$ & $0.158$ & $0.01$ \\
 $2$ & $0.41$ & $0.265$ & $0.767$ & $0.150$ & $0.01$ \\
 $3$ & $0.08$ & $0.280$ & $0.745$ & $0.162$ & $0.01$ \\
\hline
\end{tabular}
\end{center}
\end{table}

Loci of the same colors that are superimposed on the
cluster giants represent evolutionary tracks for 0.787, 0.767, and $0.745 \msol$,
which are the adopted initial masses for the same three helium abundances, in
the direction of increasing $Y_i$, for which the predicted age at the RGB tip is
approximately 12.8 Gyr.  These tracks provide a somewhat better fit to the giant
branch than the isochrones that are plotted in the previous figure, in part 
because of small differences in the treatment of the surface boundary condition
in the MESA and Victoria codes (as already noted in \S\,\ref{sec:theory}), but
also because the isochrones in Fig.~\ref{fig:bvm18} had been adjusted to the red
by $\delta(B-V) = 0.011$\ mag in order to match the observed TO color.
 
The observational data in the right-hand panels are shown with Poisson error
bars.  The outlying point in the upper panel corresponds to a group of
observed HB stars with $M_V\approx 0.7$ mag that our models cannot fully explain
because they spread below the ZAHB.  The mean masses that are lost by the RGB
stars in M\,55 are estimated to be $\Delta{\cal M}_i\approx 0.157\,{\cal
M}_\odot$ for ${\cal M}_i\approx 0.766\,{\cal M}_\odot$, which is just slightly
less than the amount expected from the \citet{rei75} formula with the parameter
$\eta_\mathrm{R} = 0.45$ (see Fig.~5 in Paper~II).  According to our simulations
of the reddest HB stars in M\,55, most if not all of its RR Lyrae variables are
expected to have $Y \approx 0.25$.

\subsection{The Periods of the RR Lyrae in M\,55}

\begin{figure}[t]
\plotone{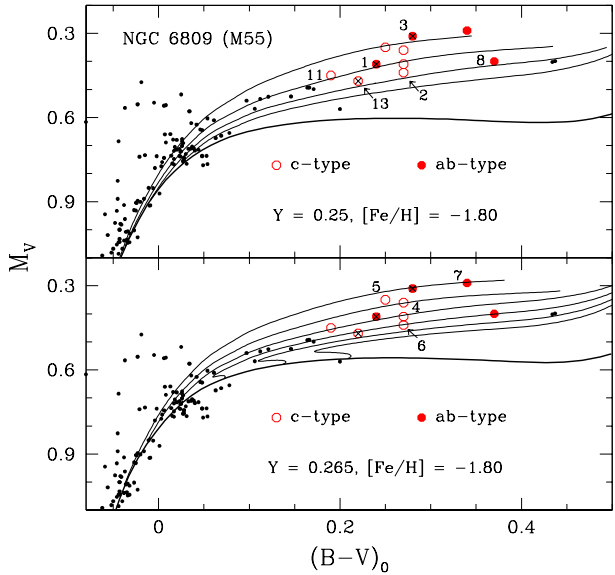}
\caption{{\it Top panel}: Fit of a ZAHB for the indicated chemical abundances
and four post-ZAHB tracks (for $0.63 \le {\cal M}/{{\cal M}_\odot} \le 0.66$,
in the direction from left to right) to M\,55 RR Lyrae (open and filled circles
in red) and non-variable blue HB stars (small black filled circles), assuming
$E(B-V) = 0.120$ and $(m-M)_V = 13.97$.  The symbols for variables V1, V3, and
V13 have been superimposed by crosses to indicate that there are large
discrepancies between the predicted and observed periods (see the text).
{\it Bottom panel}: As in the upper panel, except that the models assume
$Y=0.265$ and the tracks are for masses $0.64 \le {\cal M}/{{\cal M}_\odot} \le
0.69$ (in the direction from left to right).}
\label{fig:hb2527}
\end{figure}

Figure~\ref{fig:hb2527} shows the superposition of ZAHB loci and selected HB
tracks from the grids discussed above for $Y=0.25$ (upper panel) and $Y=0.265$
(lower panel) onto the HB of M\,55, assuming $E(B-V) = 0.120$ and $(m-M)_V =
13.97$.  The filled and open circles (in red) represent the RRab and RRc
variables, respectively.  Once the
luminosities, effective temperatures, and masses at these CMD locations have
been determined using linear interpolations within, or minor linear
extrapolations from, the tracks, the predicted periods of the RR Lyrae follow
from equations (1) and (2).  (The relevant values of $Z$ are
$4.290\times 10^{-4}$, if $Y=0.25$, and $4.203\times 10^{-4}$, if $Y=0.265$.)
 
The differences between the predicted and observed periods (in days) of the 10
RR Lyrae are listed in Table~\ref{tab:t3}, which also contains the results that
are obtained if the same models are fitted to the observations, but a smaller
value of $(m-M)_V$ by 0.05 mag, or a larger value by 0.03 mag is adopted.  In
both of the latter cases, the reddening was set (to the indicated values) so
that the models always provide essentially the same fit to the blue HB stars
with $M_V \gta 0.8$.  Note that a reduced distance modulus by 0.05 mag would
imply an increased turnoff age by about 0.5 Gyr, and vice versa (see VBLC13, their Fig.~2).  
(Because the plots for the additional values of $E(B-V)$ and $(m-M)_V$ that we
have considered in producing Table~\ref{tab:t3} are quite similar to those given
in Figs.~\ref{fig:bvm18} and \ref{fig:hb2527}, we have opted not to include
them in this paper.  Indeed, there is sufficient ambiguity in the fits of ZAHB
models and isochrones to the observed CMD of M\,55 that they cannot be used to
discriminate between the three cases considered in Table~\ref{tab:t3}.  However,
it is clear from the tabulated results for the individual stars and the values
of $\langle\delta P\rangle$ that the predicted periods are quite sensitive to
the adopted cluster properties.)

\begin{table*}[ht]
\begin{center}
\caption{Predicted minus Observed Periods}
\label{tab:t3}
\begin{tabular}{cccccccccc}
\hline
\noalign{\vskip 2pt}
\hline
\noalign{\vskip 1pt}
 & & \multicolumn{2}{c}{$(m-M)_V=13.92$} & & \multicolumn{2}{c}{$(m-M)_V=13.97$} &
 & \multicolumn{2}{c}{$(m-M)_V=14.00$} \\
 & & \multicolumn{2}{c}{$E(B-V)=0.130$} & & \multicolumn{2}{c}{$E(B-V)=0.120$} &
 & \multicolumn{2}{c}{$E(B-V)=0.116$} \\
Var. & & $Y=0.25$ & $Y=0.265$ & & $Y=0.25$ & $Y=0.265$ & & $Y=0.25$ & $Y=0.265$ \\
\hline
\noalign{\vskip 1pt}
\multicolumn{10}{c}{$ab$-type} \\ [0.5ex]
 V1 & & $-0.128$\phantom{*} & $-0.139$\phantom{*} & & $-0.089$\phantom{*} &
      $-0.098$\phantom{*} & & $-0.070$\phantom{*} & $-0.078$\phantom{*} \\
 V3 & & $-0.093$\phantom{*} & $-0.102$\phantom{*} & & $-0.044$\phantom{*} &
      $-0.053$\phantom{*} & & $-0.019$\phantom{*} & $-0.028$\phantom{*} \\
 V7 & & $+0.023$\phantom{*} & $+0.011$* & & $+0.085$\phantom{*} & $+0.071$* & &
      $+0.117$\phantom{*} & $+0.101$* \\
 V8 & & $-0.027$* & $-0.046$\phantom{*} & & $+0.031$\phantom{*} & $+0.016$* & &
      $+0.061$\phantom{*} & $+0.047$* \\
\noalign{\vskip 1pt}
\multicolumn{10}{c}{$c$-type} \\ [0.5ex]
 V2 & & $-0.019$* & $-0.031$\phantom{*} & & $+0.015$\phantom{*} & $+0.006$* & &
      $+0.031$\phantom{*} & $+0.023$* \\
 V4 & & $+0.035$\phantom{*} & $+0.027$* & & $+0.070$\phantom{*} & $+0.063$* & &
      $+0.087$\phantom{*} & $+0.081$* \\
 V5 & & $+0.020$\phantom{*} & $+0.014$* & & $+0.053$\phantom{*} & $+0.047$* & &
      $+0.069$\phantom{*} & $+0.064$* \\
 V6 & & $+0.011$\phantom{*} & $+0.001$* & & $+0.044$\phantom{*} & $+0.036$* & &
      $+0.061$\phantom{*} & $+0.054$* \\
V11 & & $-0.010$* & $-0.018$\phantom{*} & & $+0.015$\phantom{*} & $+0.008$* & &
      $+0.027$\phantom{*} & $+0.021$* \\
V13 & & $-0.078$\phantom{*} & $-0.089$\phantom{*} & & $-0.050$\phantom{*} &
      $-0.058$\phantom{*} & & $-0.036$\phantom{*} & $-0.043$\phantom{*} \\
\noalign{\vskip 3pt}
  & & \multicolumn{2}{c}{$\langle\delta P\rangle = -0.004\pm 0.025$} & &
     \multicolumn{2}{c}{$\langle\delta P\rangle = +0.035\pm 0.026$} & &
     \multicolumn{2}{c}{$\langle\delta P\rangle = +0.056\pm 0.029$} \\
\noalign{\vskip 1pt}
\hline
\noalign{\vskip 1pt}
\multicolumn{10}{l}{Asterisks indicate the data used in calculating
$\langle\delta P\rangle$ and the associated $1\,\sigma$\ uncertainties.}
\end{tabular}
\end{center}
\end{table*}

As already mentioned, the measured periods of V1, V3, and V13 are difficult to
explain insofar as they are much higher than the periods that are inferred from
the CMD locations of these stars --- as indicated by the large, negative values
of $\delta P$ in Table~\ref{tab:t3}.  Because we do not have similar problems
with the other RR Lyrae, crosses have been superimposed on the symbols representing
these variables in Fig.~\ref{fig:hb2527} to indicate that they have been
dropped from the remainder of our analysis.  However, it is odd that V1 is bluer
than most of the $c$-type variables ($\langle P_c\rangle \approx 0.38$\ d),
which is inconsistent with its high period (0.5800 d).  V3 similarly seems
anomalous in having quite a high period (0.6620 d) despite having nearly the
same color, and hence $\teff$, as the reddest first overtone pulsators, which
have shorter periods by $\sim 0.26$ d.  (V3 is more luminous, but the luminosity
difference is far too small to explain such a large discrepancy.)  Finally, the
similarity of the periods of V13 and V2, which have similar luminosities, is at
odds with the significant difference in their colors (and presumably their
temperatures).  It goes without saying that further work is needed to resolve
such apparent inconsistencies, which may be the manifestation of
deficiencies in our understanding of pulsation physics, though we suspect that
they are most likely due to errors in the derived mean properties if, indeed,
V1, V3, and V13 are cluster members.  (The periods of these stars cannot be the
problem since they are determined to very high accuracy and precision from the
light curves.)

Table~\ref{tab:t3} indicates that our models provide the best fits to the
periods of the other RR Lyrae if M\,55 has $(m-M)_V \approx 13.92$ (and $E(B-V)
\approx 0.130$). (In all three cases that are tabulated, asterisks have been
attached to the smallest $\delta P$ values.)   A modulus as high as $(m-M)_V =
14.00$ appears to be ruled out, given that the $\delta P$ values for most of
the variables are unacceptably high for any $Y \le 0.265$; as indicated at
the bottom of the table, $\langle P\rangle = 0.056 \pm 0.029$ if all of them
have $Y=0.265$.  Even higher $Y$ is unlikely because the tracks for many of the
variables would then have long blue loops, and the predicted ZAHB locations of
many of the variables would be inside the instability strip.  Since there are no
RR Lyrae at such faint luminosities, it is much more credible that they
originated from initial structures on the blue side of the instability strip
where nearly all of the non-variable HB stars are found.  Interestingly, if
$(m-M)_V \approx 13.92$, the models for $Y=0.25$ predict periods that are closer to
the observed periods of V2, V8, and V11 than those for $Y=0.265$, whereas the
higher $Y$ models are favored in the case of V4--V7.  However, such results are
no more than suggestive because a change in $Y$ mainly affects the predicted
mass at a given CMD location, which has a relatively minor impact on the
pulsation period (see equations 1 and 2).  According to the results presented
in Table~\ref{tab:t3}, an increase in the helium abundance by $\delta Y = 0.015$
implies a reduced period by only $\sim 0.005$--0.015\ days (if the
temperature and luminosity are fixed).

Previous studies (VBLC13, Papers I and II) have not found any indications of
substantial errors in the temperatures of our models for the HB
phase, as they are able to reproduce the detailed morphologies of observed HBs
over wide ranges in color and luminosity very well.  (The same can be said of
stellar models applicable to the TO and upper MS stars in GCs and in the field;
see \citet{bsv10} and \citet[2014a,b]{vcs10}, who show that the model and IRFM
$\teff$\ scales are very similar and that isochrones generally provide quite
consistent interpretations of observations on many different color-magnitude
planes.)  The predicted temperatures for warmer stars, such as those found in
the instability strip and along the blue HB, should be particularly robust
because surface convection zones are very thin or absent in them.  Lacking any
compelling evidence to the contrary, we are inclined to believe that the model
$\teff$\ scale is accurate to within $\sim\pm 80$~K ($1\,\sigma$), which is
comparable to the uncertainties associated with empirical IRFM-based
temperatures (e.g., see CRMBA).

For a typical RR Lyrae variable with $\teff = 6750$~K ($\log\teff = 3.83$), an
error of $\pm 80$~K in its temperature would translate to an error in $\log P$
amounting to $\pm 0.017$ (see equations 1 and 2), which corresponds to
$\delta P = 0.016$~d if the star's pulsation period is 0.400~d, or 0.024~d if
$P = 0.600$~d.  Our HB tracks are based on up-to-date physics (see Paper I),
including a treatment of mixing at the boundary of a convective He-burning core
that is supported by asteroseismic studies of field HB stars (see \citealt[and
references therein]{ccc15}); consequently, they should provide especially
realistic predictions of the paths that HB stars follow on the H-R diagram.  In
any case, errors in $\log (L/L_\odot)$ (as well as in $\log ({\cal M}/{{\cal
M}_\odot})$ and $Z$) have significantly smaller effects on the predicted periods
of RR Lyrae than errors in $\log\teff$.  Based on these considerations, we
expect that our models should be able to reproduce measured periods to within
$\sim 0.03$~d.  (This should be quite a realistic estimate since we have shown
in Paper I that both the slope and the zero-point of the best available
empirical $M_V$ vs.~[Fe/H] relation for RR Lyrae stars is well reproduced by our
models.  This issue is revisited in \S\,\ref{sec:candle}.)

Additional fits of our models to the HB of M\,55 (not shown) indicate that the
cluster must have $(m-M)_V \gta 13.88$ in order to provide an acceptable
interpretation of both the pulsational properties of the cluster RR Lyrae and
the non-variable HB stars just blueward of the instability strip.  Otherwise,
the latter would lie fainter than the ZAHB and we would obtain $\langle\delta
P\rangle < -0.03$~d for the variable stars.  Even shorter distances would
run into the aditional difficulties of requiring $E(B-V) > 0.140$, for which
there is no observational support, and the predicted TO age would be similar to,
or exceed, the age of the universe ($\approx 13.8$ Gyr; \citealt{pc15},
\citealt{blw13}).  As a result of these considerations, we conclude that M\,55
has $(m-M)_V = 13.93 \pm 0.05$.  (This assumes, of course, that our evolutionary
computations for the HB phase, and equations (1) and (2), accurately predict the
properties of real stars in the core He-burning phase.)  Due to the reduced
distance modulus, the age derived previously ($\approx 12.7$ Gyr, see
Fig.~\ref{fig:bvm18}) rises to $\approx 13.1$ Gyr. 

The apparent distance modulus that is derived in this way cannot be very 
dependent on the adopted metallicity because
almost the same reddening is required to fit the blue HB stars just below the
knee (this will be true even if the adopted [Fe/H] value is changed by as much
as 0.3--0.4 dex; recall Fig.~\ref{fig:hbmet}), and consequently, the
temperatures that are derived for the RR Lyrae will be very similar.  Hence, if
models for different [Fe/H] values (within some reasonable range, say 0.25
dex) are to predict close to the measured periods, the luminosities of the
variable stars cannot have much of a dependence on metallicity either.
(Changing only the $M_V$ values of the RR Lyrae by, e.g., $-0.04$ mag,
which is approximately the vertical shift of a ZAHB for [Fe/H] $= -2.0$
relative to one for [Fe/H] $= -1.8$, at the color of the instability strip,
would alter the predicted periods by $\delta\log P\approx 0.014$, or
$\delta P\approx 0.019$~d if $P = 0.600$~d.)  Some differences in the predicted
masses and in the value of $Z$ corresponding to the adopted [Fe/H] value can be
expected, but pulsation periods have considerably less sensitivity to these
quantities than to $\log(L/L_\odot)$ or $\log\teff$\ (see equations 1 and 2).

While the uncertainties associated with the distance moduli as derived
from our HB simulations, on the one hand, and from the RR Lyrae, on the other,
overlap one another if [Fe/H] $\approx -1.8$, only marginal consistency would
have been obtained had we adopted [Fe/H] $= -2.0$.  As noted in the previous
paragraph, fits of HB models for the lower metallicity to the photometric
observations would yield $(m-M)_V \approx 14.01$, which is 0.08--0.09 mag larger
than the distance modulus at which the predicted mean period of the RR Lyrae,
$\langle P\rangle$, would be in good agreement with the observed value.  Since
this difference is approximately double that which is obtained if M\,55 has
[Fe/H] $= -1.8$, the RR Lyrae indicate a clear preference for [Fe/H] $\gta -1.8$
over a lower metallicity.  We will now consider the additional constraint
that is provided by a member binary star with well determined properties.


\subsection{The Eclipsing Binary V54 in M\,55}
\label{subsec:v54}

Having derived the distance modulus of M\,55 to within $\sim\pm 0.05$\ mag 
($1\,\sigma$) from its RR Lyrae and non-variable HB stars, assuming that our
stellar models for the core He-burning phase are reliable, we can now use the
properties of the detached, eclipsing binary in this cluster, V54, to constrain
the effective temperatures and the chemical composition of its components.  For
this part of the analysis, we need only those fundamental properties of V54
that are listed in Table~\ref{tab:t4}, which have been taken from the study by
\citet{ktd14}.  From the measured $V$ magnitudes and their uncertainties, it
follows that the primary and secondary components of the binary have $M_V =
4.47\pm 0.06$ and $7.05\pm 0.07$, respectively, if the apparent distance modulus
of the cluster is $(m-M)_V = 13.93\pm 0.05$.

\begin{table}[b]
\begin{center}
\caption{Basic Properties of V54}
\label{tab:t4}
\begin{tabular}{lcc}
\hline
\noalign{\vskip 2pt}
\hline
\noalign{\vskip 1pt}
Property & Primary & Secondary \\
\hline
\noalign{\vskip 1pt}
 Mass ($\cal M_\odot$) & $0.726 \pm 0.015$           & $0.555\pm 0.008$ \\
 Radius ($R_\odot$)    & $1.006 \pm 0.009$           & $0.528\pm 0.005$ \\
 $B$ magnitude         & $18.93 \pm 0.01\phantom{0}$ & $21.87\pm 0.03\phantom{0}$ \\
 $V$ magnitude         & $18.40 \pm 0.01\phantom{0}$ & $20.98\pm 0.02\phantom{0}$ \\
\noalign{\vskip 1pt}
\hline
\end{tabular}
\end{center}
\end{table}

\begin{figure}[t]
\plotone{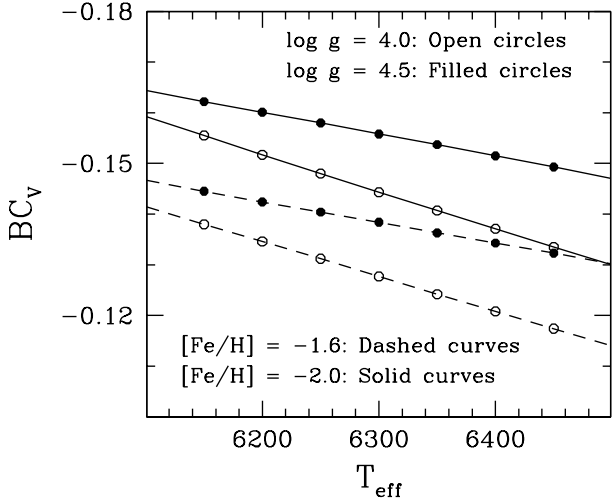}
\caption{The $\teff$\ dependence of the bolometric corrections in the $V$ band
for stars with [Fe/H] $= -1.6$ and $-2.0$ assuming, in each case, $\log\,g =
4.0$ and 4.5 (as indicated).  Turnoff stars in M\,55 are predicted to have
temperatures near 6300~K.}
\label{fig:bcv}
\end{figure}

The extinction, $A_V$, is not needed to compute these $M_V$ values because the
absolute magnitude scale was set by our HB models, though the apparent modulus
that was derived does depend to some extent (see Table~\ref{tab:t3}) on the
adopted reddening.  To convert the $M_V$ values to absolute bolometric
magnitudes, it is necessary to know the temperatures of the components, which
can be derived from the dereddened $B-V$ colors.  However, the bolometric 
corrections in the $V$ band ($BC_V$) that apply to metal-poor stars located
near the turnoffs of GCs are very weak functions of $\teff$, so it does not
matter if the temperatures that are assumed for this part of the analysis are 
not accurate.  As shown in Figure~\ref{fig:bcv}, which considers metallicities
in the range $-2.0 \le$ [Fe/H] $\le -1.6$, the $BC_V$ values of upper MS and TO
stars vary, at a fixed value of $\log\,g$, by only $\sim 0.03$ mag over a 400~K
range in $\teff$.  (These results were obtained from the transformations
provided by \citet{cv14}, who computed $BC$s based on MARCS model atmospheres
(\citealt{gee08}) for many filter passbands over wide ranges in [Fe/H],
[$\alpha$/Fe], $\log\,g$, and $\teff$.)

In any case, it is easy to obtain consistency between the temperatures that
are derived from the luminosities and radii of the components of V54 and those
assumed in the determination of the $BC_V$ values simply by iterating between
the two.  To be more explicit: we adopted initial values of $\teff$\ that were
calculated from the IRFM-based $(B-V)_0$--$\teff$--[Fe/H] relation of CRMBA (see
their Table 4), determined the corresponding values of $BC_V$ from the
transformation tables provided by \citet{cv14}, converted the resultant
$M_{\rm bol}$ values to $\log\,(L/L_\odot)$, and then calculated the temperatures
of the binary from those luminosities and the measured radii.  These $\teff$
values could then be used to recalculate the bolometric corrections, etc.  After
4 iterations of this procedure, the input and output temperatures were the
same, resulting in $\teff = 6347\pm 116$~K and $5009\pm 104$~K, in turn, for
the primary and secondary of V54.  (The error bars were calculated from the
$1\,\sigma$ uncertainties in the luminosities and radii.)  These estimates
compare very well with the values of $\teff = 6361$~K and 5050~K, respectively,
that are obtained from the CRMBA color-temperature-metallicity
relation.  Worth emphasizing is that our temperature determinations are
independent of those inferred from spectra (e.g., the fitting of Balmer line
profiles) or from the application of the IRFM.

\begin{figure}[t]
\plotone{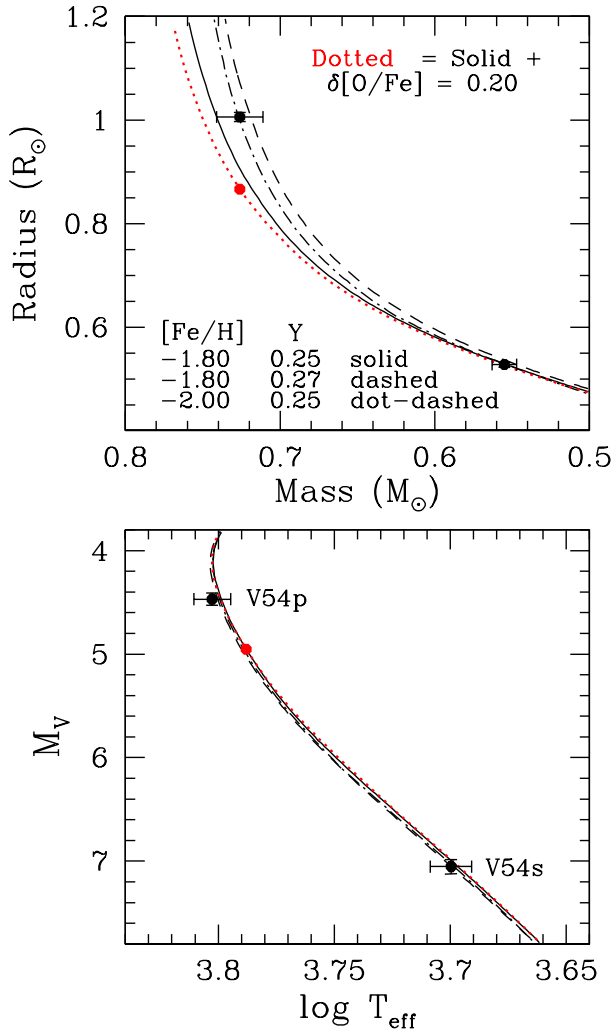}
\caption{{\it Top panel}: Comparison of the measured masses and radii of the
components of V54 with the mass-radius relations predicted by isochrones for
the indicated chemical abundances that provide the best fits to the turnoff
photometry of M\,55 (see the text).  The large filled circle in red indicates
the location of a stellar model along the dotted red isochrone that has the
same mass as the primary of V54.  {\it Bottom panel}: as in the top panel,
except that the comparisons are made on the ($\log\teff,\,M_V$)-diagram.}
\label{fig:mrhr}
\end{figure}

As shown in Figure~\ref{fig:mrhr}, the isochrones that provide the best fits to
the observed CMD of M\,55, on the assumption of the various chemical abundances
that are specified in the lower left-hand corner of the upper panel, provide an
excellent fit to the binary components on both the mass--radius plane and the
$(\log\teff,\,M_V)$-diagram.  Recall that the isochrone for [Fe/H] $= -1.80$
[$\alpha$/Fe] $= +0.4$, and $Y=0.25$ has an age of 13.1 Gyr.  This rises to
$\approx 14.0$\ Gyr if the cluster turnoff is fitted by an isochrone for
[Fe/H] $= -2.0$, assuming the same values of [$\alpha$/Fe] and $Y$.
For the higher metallicity case, a 13.1 Gyr isochrone with $Y = 0.27$ has also
been plotted (the dashed curve) to illustrate the impact of varying this
parameter.  (M\,55 probably has stars with even higher helium abundances
given that such stars are expected to produce the bluest HB stars; see the
discussion in Paper II of the especially long blue HB tail in M\,13.  In
principle, this should be taken into account when fitting isochrones to the CMD
of M\,55, but this would cause only a minor perturbation to the derived age.)

The effect of increasing the oxygen abundance by 0.2 dex is shown by the dotted
curve (in red).  In this 12.6 Gyr isochrone, which provides as good a fit to
the TO photometry as the other isochrones (on the assumption of the same
distance modulus), the abundances of all of the other elements are the same as
those assumed in the solid curve.  The large red filled circle shows where a
model for the same mass as the primary of V54 sits on the dotted isochrone in
the two panels of Fig.~\ref{fig:mrhr}.

While the various isochrones are nearly coincident on the $\log\teff$--$M_V$
plane, they are quite well separated on the mass--radius diagram, which 
provides a much better discriminant of the assumed chemical abundances.  
Unfortunately, even though the mass of the primary is known to within 2.1\% (see
Table~\ref{tab:t4}), this uncertainty is still too large to place really tight
constraints on $Y$.  If M\,55 has [Fe/H] $= -1.80$, [$\alpha$/Fe] $= 0.4$,
$(m-M)_V = 13.93$, and an age of $\approx 13.1$ Gyr, the predicted masses for
any helium abundance in the range $0.25 \lta Y \lta 0.28$ are consistent with
the measured mass of the primary to within its $1\,\sigma$ error bar.  The upper
panel of Fig.~\ref{fig:mrhr} also indicates a preference for [O/Fe] $\le 0.4$
if [Fe/H] $= -1.80$, though isochrones for a higher oxygen abundance by as much
as $\sim 0.3$ dex (estimated by extrapolating the separation between the solid
and dotted curves at the observed radius and then applying the result to the
dot-dashed curve) would satisfy the mass constraint if M\,55 has [Fe/H] $\approx
-2.00$.  Indeed, a higher oxygen abundance would be needed in this case to avoid
a conflict between the age of M\,55 and the age of the universe.  Thus, [Fe/H]
$< -2.0$ can be ruled out if M\,55 has [O/Fe] $\lta 0.4$.  Models for [Fe/H] $>
-1.80$ could be accommodated if they have [O/Fe] $< 0.4$, which would, however,
increase the predicted age at a given value of $(m-M)_V$.  Based on these
considerations, it would appear that M\,55 has an age between $\approx 13.0$
Gyr and 13.8 Gyr, where the upper limit is set by the age of the universe.

However, all of these inferences assume that V54 has $Y\approx 0.25$.  Indeed,
some of the stars in M\,55 probably do have such helium abundances, but others
are likely to have higher $Y$.  This is suggested by our analysis of the cluster
RR Lyrae, but more importantly, it is becoming clear that a spread in $Y$ is a
common characteristic of GCs (see, e.g., \citealt{pba07};
\citealt[\citealt{mmp13}]{mmp12}; \citealt{nmp15}, and our examination of both
the variable and non-variable stars in M\,3, M\,13, and 47 Tuc that were
presented in Paper II).  Hence, it is easily possible that V54 is a member of
a helium-enhanced population in M\,55, in which case, models for [Fe/H] values
as high as $\sim -1.6$, or those for, e.g., [Fe/H] $= -1.8$ and [O/Fe] $= 0.6$,
could satisfy the binary constraint if $Y \lta 0.28$ (an estimate based on
the results that are shown in the upper panel of Fig.~\ref{fig:mrhr}).  Since
[O/Fe] $=0.6$ (i.e., an enhancement of 0.2 dex above the amount implied by
[$\alpha$/Fe] $= 0.4$) is a viable possibility, the absolute age of M\,55 could
be anywhere in the range from $\sim 12.6$ to 13.8 Gyr.

\begin{figure}[t]
\plotone{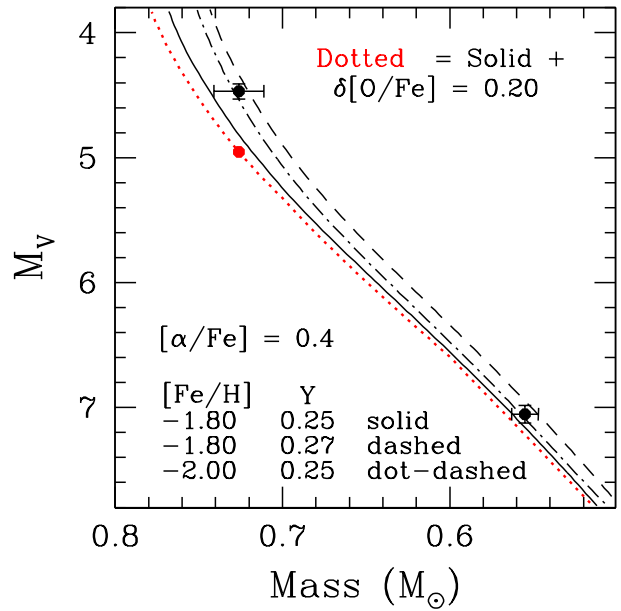}
\caption{As in the previous figure, except that the comparisons are made on the
mass--$M_V$ diagram.}
\label{fig:mmv}
\end{figure}

The mass--$M_V$ diagram, which is given in Figure~\ref{fig:mmv}, looks
qualitatively very similar to the mass--radius diagram, and its implications
for the cluster properties are clearly nearly the same as those just described.
The main strength of stellar models has always been the prediction of the
luminosities of stars; consequently, Fig.~\ref{fig:mmv} provides a much more
compelling comparison between theory and observations than those shown in the
previous figure, though our success in matching the radii and temperatures of
the V54 components as well as their luminosities, is very encouraging.  Even
though we are unable to place very tight limits on the chemical abundances of
M\,55 from comparisons of predicted mass-luminosity relations with the
properties of the binary V54, it is comforting that the results from what is
effectively a {\it stellar interiors} approach are consistent with, while
being independent of, spectroscopically derived abundances.

In view of the high age that we have found for M\,55 on the assumption of
$(m-M)_V = 13.93$, a reduced distance modulus by more than $\sim 0.05$ mag is 
unlikely because the consequent cluster age would be within $\sim 0.5$ Gyr of
the age of the universe even if M\,55 has [O/Fe] $\approx 0.6$ (assuming
$-1.8 \lta$ [Fe/H] $\lta -2.0$).  This provides an indirect argument that the
periods of the cluster RR Lyrae that are inferred from our HB tracks cannot be
less than the observed periods by more than 0.03 d.  What are the consequences,
then, of adopting a larger distance modulus by 0.05 mag (i.e., $(m-M)_V =
13.98$), in which case the predicted RR Lyrae periods would be {\it greater}
than the observed periods by $\sim 0.03$~d?

For one thing, an increased distance would make V54 intrinsically brighter and
its components would also be hotter, since their temperatures are calculated
directly from their radii and luminosities.  If $(m-M)_V = 13.98$, we obtain
$\teff = 6415$~K and 5051~K, in turn, for the primary and secondary components.
These estimates are still within the $1\,\sigma$ error bars of the temperatures
previously derived on the assumption of the shorter distance modulus and,
importantly, they are comparable with, or higher than, the temperatures found
from the application of the IRFM by CRMBA.  Indeed, the binary in M\,55
provides compelling support for a relatively warm $\teff$\ scale at low
metallicities, which is one of the main results of this investigation.

\begin{figure}[t]
\plotone{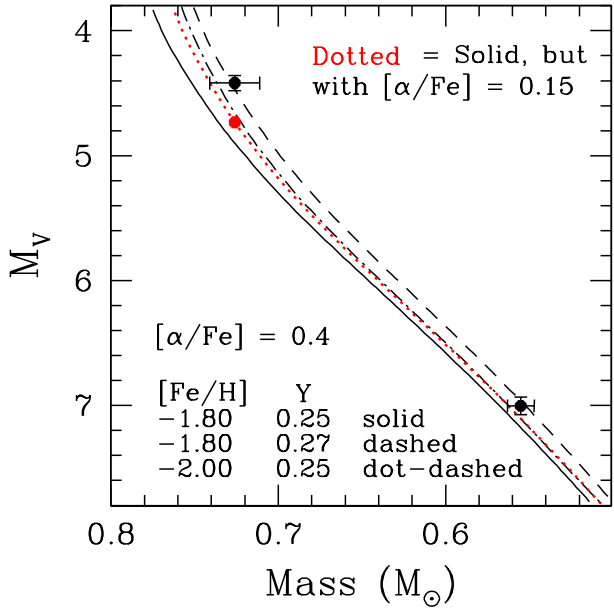}
\caption{As in the previous figure, except that a larger distance modulus by
0.05 mag has been assumed (specifically $(m-M)_V = 13.98$), and younger
isochrones by 0.5--0.6 Gyr have been plotted so as to be consistent with the
increased distance.  Note that, as a result of the change in $(m-M)_V$, the
components of V54 will be instrinically more luminous and hotter.}
\label{fig:mmvmod}
\end{figure}

Another consequence of adopting $(m-M)_V = 13.98$ is that the turnoff age would
be reduced by $\sim 0.5$--0.6 Gyr.  Because younger ages imply higher masses at
a given luminosity along the main sequence, the predicted mass--$M_V$ relations
for the same chemical abundances considered in Fig.~\ref{fig:mmv} would be
shifted somewhat to the left of their locations therein.  As shown in 
Figure~\ref{fig:mmvmod}, isochrones for $Y = 0.25$ and [Fe/H] $\sim -1.8$
would be in poorer agreement with the properties of the binary than in
Fig.~\ref{fig:mmv} due to the net effect of this shift and the revised $M_V$
values of the binary.  On the other hand, the models for $Y \gta 0.26$ or
for [Fe/H] $\sim -2.0$ and $Y \approx 0.25$ would still provide a satisfactory
fit to the data to within $1\,\sigma$.

However, if V54 has [Fe/H] $= -1.80$ and $Y=0.25$, it is still possible to
obtain consistency of the predicted and observed masses to within $1\,\sigma$
if a reduced O abundance is assumed.  This is illustrated by the dotted
isochrone (in red, from \citealt{vbf14}), which assumes [$\alpha$/Fe] $= 0.15$.  
(At low metallicities, the effects on isochrones of varying [$\alpha$/Fe] are 
due almost entirely to the associated changes in the O abundance; see
\citealt{vbd12}.)  To obtain a consistent fit to the turnoff photometry in this
case, a higher age would have to be assumed, $\approx 13.3$\ Gyr, which is not
significantly different from the age that was derived on the assumption of
$(m-M)_V = 13.93$.  On the other hand, a $\sim 12.6$ Gyr isochrone for [Fe/H]
$= -1.80$ and [$\alpha$/Fe] $= 0.4$ would provide a consistent interpretation
of all of the observations (i.e., both the cluster CMD and the properties of
V54) if the binary has $Y \approx 0.26$--0.28 (see Fig.~\ref{fig:mmvmod}).  If
the helium abundance is high enough, models for [O/Fe] $= 0.6$ would also 
satisfy the binary constraint, and the resultant turnoff age would be close to
12.0 Gyr. 

\begin{figure}[t]
\plotone{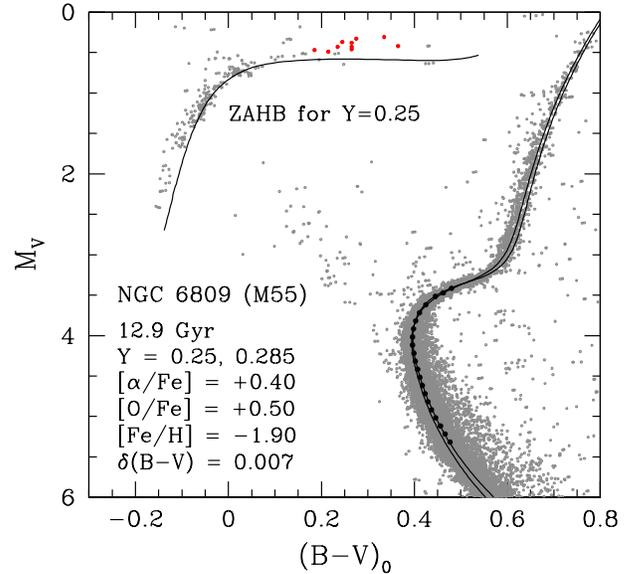}
\caption{Similar to Fig.~\ref{fig:bvm18}, except that stellar models for
different chemical abundances (as indicated) have been compared with the CMD
of M\,55.  To match the cluster HB stars to a ZAHB for $Y=0.25$,
compromise values of} $E(B-V) = 0.125$ and $(m-M)_V = 13.95$ have been
assumed (see the text).
\label{fig:m55fit}
\end{figure}

To summarize this section: our consideration of HB simulations and RR Lyrae
periods suggests that M\,55 has an apparent distance modulus in the range
$14.93 \lta (m-M)_V \lta 14.97$ and [Fe/H] $= -1.85 \pm 0.1$, which also
satisfies the constraints provided by the eclipsing binary V54 to within the
uncertainty of its helium abundance.  Assuming that the cluster has [O/Fe]
$= 0.5\pm 0.1$, with [$m$/H] $= 0.4$\ for the other $\alpha$-elements, and that
the faintest stars in the vicinity of the knee of the HB have $Y = 0.25$, our
best estimate of its age is $12.9 \pm 0.8$\ Gyr, where the error bar takes into
account the effects of the distance and chemical abundance uncertainties
(approximately $\pm 0.5$\ Gyr and $\pm 0.3$\ Gyr, respectively).  The fit of a
ZAHB and a 12.9 Gyr isochrone for these chemical abundances to the CMD of M\,55
is shown in Figure~\ref{fig:m55fit}.  For illustrative purposes, an isochrone
for the same age but for $Y=0.285$\ has also been plotted; according to our HB
simulations, few, if any, of the stars in M\,55 have higher helium abundances.

\begin{figure*}[t]
\plotone{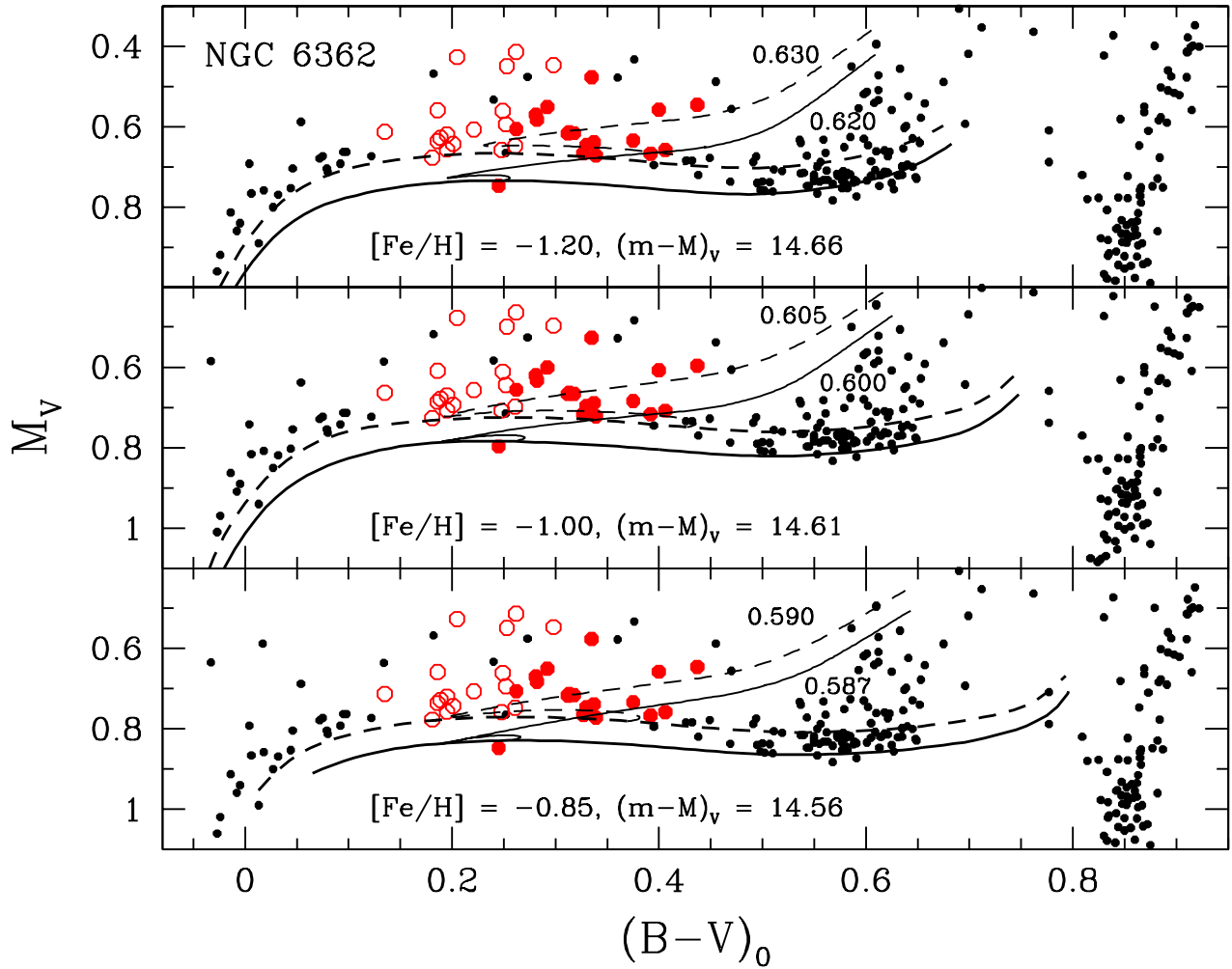}
\caption{Fits of ZAHB loci for $Y=0.25$ (thick solid curves) and $Y=0.265$ or
0.270 (thick dashed curves in the bottom or the top two panels, respectively)
for the indicated [Fe/H] values and apparent distance moduli to the non-variable
HB stars (small black filled circles) and the RR Lyrae in NGC\,6263.  The $ab$-
and $c$-type variables have been plotted, in turn, as large filled and open
circles in red.  Cluster giants in the same magnitude range are located at
$(B-V)_0 \gta 0.8$.  In all three panels, $E(B-V) = 0.070$ has been adopted (see
the text for the justification of this choice).  Two evolutionary tracks (for
the indicated masses, in solar units) have also been plotted in each panel to
illustrate their strong morphological dependence on $Y$.}
\label{fig:hb3}
\end{figure*} 

\section{NGC\,6362}
\label{sec:n6362}

The basic properties of NGC\,6362 appear to be relatively well determined.
Most estimates of the foreground reddening fall in the range $0.07 \lta E(B-V)
\lta 0.09$ (e.g., see \citealt{sfd98}, \citealt{bcr99}, \citealt{okt01},
\citealt{sf11}, and the 2010 edition of the catalogue by \citealt{har96}).
Similar good consistency has been found for the cluster metallicity over the
years, with the majority of studies finding [Fe/H] values between $-0.96$ (CG97)
and $-1.15$ (KI03), including the investigations by, e.g., \citet{zw84}, CBG09,
\citet{mdm16}, and \citet{mmd17}.  As first reported by \citet{dmb14},
Mucciarelli et al.~and Massari et al.~have confirmed that, in common with most
GCs, NGC\,6362 contains multiple, chemically distinct stellar populations.

Because NGC\,6362 has a well-populated red HB, along with a sufficient number 
of non-variable blue HB stars in the vicinity of the knee to provide a useful
constraint on the reddening, fits of ZAHB models to the observed HB
should be reasonably straightforward.  However, it turns out that a
single ZAHB locus cannot provide a satisfactory fit to the faintest HB stars
across the entire color range that they occupy.  The problem is that, as
pointed out by \citet{bcr99}, the HB of NGC\,6362 has an odd HB morphology in
that the faintest stars just to the blue of the instability strip are $\sim 0.1$
mag brighter than the faintest of the red HB stars; i.e., there is a significant
downward tilt of the HB in the direction from blue to red.  Although Brocato
et al.~suggested that variations in the bolometric corrections with [Fe/H] and
$\teff$\ may be responsible for this behavior, this speculation is not supported
by our HB models.  We believe, in fact, that the observed morphology is a
manifestation of the multiple stellar populations phenomenon.

Figure~\ref{fig:hb3} illustrates the fits of ZAHB loci for $Y = 0.25$ (the
thick solid curves) and either 0.27 (the thick dashed curves in the top and
middle panels) or 0.265 (bottom panel) to the HB of NGC\,6362 on
the assumption of three different [Fe/H] values and distance moduli that have
been derived by matching the faintest stars in the red HB to the ZAHB for
$Y=0.25$ (as specified in each panel).  Large open and filled circles, in red,
indicate the locations of the RR Lyrae, for which \citet{okt01} provide
intensity-weighted mean magnitudes and colors.  (To better represent the colors
of equivalent static stars, their $\langle B\rangle - \langle V\rangle$ values
have been corrected by the amounts given by \citealt{bcs95}.)  As it turns out,
the ZAHBs for the higher values of $Y$ provide good fits to the lower bound of
the main distribution of these pulsators, as well as the non-variable stars on
either side of the instability strip.  To obtain nearly identical matches
to the cluster stars in the color range $0.05 \lta (B-V)_0 \lta 0.50$, it was
necessary to adopt a slightly larger He enhancement in the top and middle
panels.

These fits assumed $E(B-V) = 0.070$, independently of the adopted
metal abundance.  Because the reddening has a direct impact on the $\teff$
scale of the RR Lyrae, we checked whether a consistent interpretation of {\it HST}
photometry for NGC\,6362 (\citealt{sbc07}) could be obtained on the assumption
of the same reddening and distance moduli; and indeed, our ZAHB models for all
three metallicities match the $F606W,\,F814W$ observations for the non-variable
stars on both the blue and red sides of the instability strip just as well as
in the three panels of Fig.~\ref{fig:hb3}.  Since $E(B-V) = 0.07$ is within
0.01 mag of dust map determinations (\citealt{sfd98}, \citealt{sf11}), this
estimate appears to be particularly well supported.  (Note that, because the
redward extent of a ZAHB is quite a strong function of metallicity, it would not
be possible to obtain a satisfactory fit of a ZAHB to the reddest HB stars if
[Fe/H] $< -1.20$.  Even [Fe/H] $ = -1.20$ presents some difficulties in this
regard as several of the cluster stars lie below the ZAHB for $Y=0.250$, though
this discrepancy could be the consequence of small errors in the model colors.) 

The distribution of the variable stars in NGC\,6362 provides further evidence
that they, along with bluer stars, have higher helium abundances than most of
the reddest HB stars.  (Since ZAHBs for $Y > 0.25$ pass through the reddest
stars, some of the latter could have higher $Y$.)  In each panel, tracks
for masses, in solar units, that are specified close to the ends of these
evolutionary sequences, have been plotted that intersect the red edge of the
main distribution of the $ab$-type RR Lyrae.  The dashed tracks (for $Y \ge
0.265$) have long blue loops before the direction of the evolution turns back
to the red, and curiously, they reproduce the locations of not only several of
the fundamental mode pulsators near the ZAHB, but also some of them at higher
luminosities.  That is, these tracks follow the morphology of the red edge of
the distribution of filled red circles remarkably well.  Even if the
computations for $Y=0.25$ did not suffer from the problem that the ZAHB is
significantly fainter than all of the variable stars, except one, they predict
blue loops that are too small to provide a comparable fit to the observations.

Interestingly, \citet{mdm16} have reported that the [Na/Fe] distribution along
the giant branch of NGC\,6362 is broad and bimodal, and that $\sim 82$\% of
the red HB stars are Na-poor, from which they conclude that Na-rich stars on
the RGB will populate the blue HB (though stars belonging to the latter were
not included in their observing program).  Thus, it would appear that there is
a strong correlation of the Na and He abundances along the HB, which would not
be at all surprising since H-burning nucleosynthesis at sufficiently high
temperatures will tend to increase the abundance of sodium, thereby causing the
O--Na anticorrelation that is a common characteristic of GCs.  Based on the
results shown in Fig.~\ref{fig:hb3}, we will initially assume that the RR Lyrae
in NGC\,6362 have $Y=0.265$ or 0.270, depending on the adopted metallicity, when
we predict their periods from their CMD locations, though the brightest and
bluest ones probably have even higher helium abundances.

\subsection{The RR Lyrae Variables in NGC\,6362}
\label{subsec:rrl6362}

The binary mass-luminosity relation (to be discussed in \S\,\ref{subsec:bin6362})
appears to favor a relatively high metallicity for NGC\,6362; consequently, we
begin our analysis of the cluster RR Lyrae by fitting ZAHB models and HB tracks
for $Y=0.265$ and [Fe/H] $= -0.85$ to the observations.  According to the bottom
panel of Fig.~\ref{fig:hb3}, most of the variable stars lie on or above this
ZAHB, though it can be expected that some fraction of the brighter stars have
somewhat greater helium abundances.  Fortunately, the helium abundance
uncertainty does not represent a serious concern for our results (see Papers I
and II) because the effects of small changes in $Y$ on the mass, and hence the
period, at a given CMD location are quite minor.

The faintest $ab$-type variable, V25, is considerably fainter than the others,
and even though its CMD location suggests that it may have a helium abundance
close to $Y=0.25$, the period predicted by models for this value of $Y$ is
smaller than the observed period by $\sim 0.06$\ days.  Such a large discrepancy
can hardly be due to a problem with our models because they are able to explain
the periods of most of the cluster variables to within $\sim 0.02$\ d (see
below), including that of the bluest of the remaining fundamental mode pulsators
(V3), which has nearly the same period and color as V25, despite being brighter
by 0.14 mag.  Such a large luminosity difference should give rise to a
difference in period of nearly 0.05~d between V3 and V25.  Because additional
work is needed to understand its anomalous properties, V25 has been dropped from
further consideration.

\begin{figure}[t]
\plotone{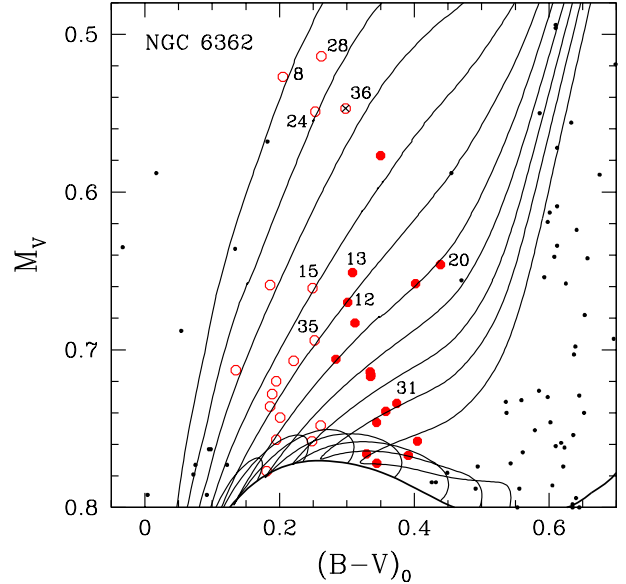}
\caption{Overlay of a ZAHB for $Y=0.265$, [$\alpha$/Fe] $= 0.4$, and [Fe/H] $=
-0.85$, along with HB tracks for 10 masses in the range $0.572 \le
{\cal M}/\msol \le 0.600$ (in the direction from left to right), onto the HB of
NGC\,6362, assuming $E(B-V) = 0.070$ and $(m-M)_V = 13.56$.  (The plot has been
stretched in the vertical direction for the sake of clarity.)  The open and
filled circles, in red, indicate the CMD locations of the $c$-type and $ab$-type
RR Lyrae, respectively.  Variables are identified by their ``V" numbers if the
predicted and observed periods differ by $> 0.030$~d.  The symbol representing 
V36 has been superimposed by a cross to indicate that it has not been included
in our analysis (see the text).}
\label{fig:hb085}
\end{figure}

The same ZAHB (for $Y=0.265$) is shown in Figure~\ref{fig:hb085} along with 9
evolutionary tracks for masses (in the direction from left to right) that range
from 0.572 to 0.596 $\msol$, in $0.003 \msol$\ increments, and a tenth track
for a mass of $0.600 \msol$.  As these tracks follow blue loops that lie very
close to the ZAHB, the plot has been stretched by a large amount in the vertical
direction so that they can be easily distinguished.  Since the tracks overlap
one another near the ZAHB, there is obviously some ambiguity in determining the
masses of the RR Lyrae in this region of the CMD.  However, this uncertainty has
only minor consequences for the predicted periods of these variables because the
range in possible masses that could apply to a given star is small.  Similarly,
differences between the assumed and actual helium abundances at the level of
$\Delta Y\lta 0.01$ will not affect the interpolated masses and predicted
periods of the RR Lyrae by very much.  For the 5 variables in the overlap zone
just above the ZAHB, masses were assigned (from the range of possible values
implied by the superposition of the tracks onto the observed stars) that
produced the best agreement between the predicted and observed periods.

Aside from the mass determination, it is straightforward to interpolate in, or
extrapolate from, the tracks to obtain the luminosities and temperatures of the
variable stars.  Since the adopted chemical abundances correspond to $Z =
3.729\times 10^{-3}$, equations (1) and (2) can be used to predict the periods
of the RR Lyrae, and it turns out that the periods so obtained are generally in
very good agreement with the observed periods.  Only for the 10 variables that
are identified in Fig.~\ref{fig:hb085} did we find differences between the
predicted and observed periods $> 0.03$\ days.  The largest discrepancy was
found for V36 ($> 0.11$~d), which was dropped from our analysis because it
clearly has anomalous properties.  This leaves us with a sample of 17 RRab and
16 RRc stars.\footnote{We have the impression from the work carried out so far
in this series of papers that the periods of variables located near the red or
blue edges of the instability strip (e.g., V20) or at the boundary between RRab
and RRc variables (e.g., V12, V13, V15, and V35) tend to be the most difficult
ones to reproduce theoretically; also see, e.g., Fig.~7 in Paper I and our study
of M\,13 in Paper II. (Since periods are well determined quantities, we
suspect that the difficulty is associated with the mean magnitudes and colors.)
It would be worth checking whether this behavior is common to the variable star
populations of most GCs since such tendencies could have important implications
for our understanding of the evolution of the pulsational properties of RR Lyrae
when they move into or out of the instability strip or when a transition is made
from fundamental to first overtone pulsation, and vice versa.}

\begin{figure}[t]
\plotone{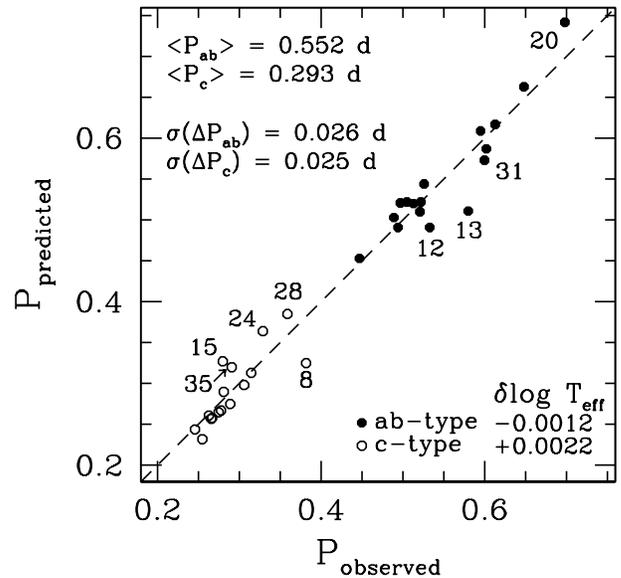}
\caption{Comparison of the observed periods, in days, of the $ab$-type and
$c$-type RR Lyrae in NGC\,6362 with those determined from models for $Y=0.265$,
[$\alpha$/Fe] $= 0.4$, and [Fe/H] $= -0.85$.  If the inferred temperatures of
the variables are adjusted by the amounts specified in the lower right-hand
corner, the mean values of $P_{ab}$ and $P_c$ from the models agree with the
observed values of $\langle P_{ab}\rangle$ and $\langle P_c\rangle$, which are
given in the top left-hand corner, to three decimal places.  The differences
between the predicted and observed periods have standard deviations amounting
to $\sigma = 0.026$ and 0.025 days (as indicated) for the RRab and RRc stars,
respectively.  Variable identification numbers are used to highlight the
locations of the outliers in this plot (see the text for some discussion of
these stars).}
\label{fig:per085}
\end{figure}

Figure~\ref{fig:per085} shows how well the observed periods of these RR Lyrae
are reproduced by our models. From the differences between the observed
periods and those calculated from equations (1) and (2), we obtain the mean
offsets $\langle\Delta P_{ab}\rangle = -0.006 \pm 0.026$~d and $\langle\Delta
P_c\rangle = 0.005 \pm 0.025$~d, where the uncertainties represent the standard
deviations of the mean.  By applying the small zero-point adjustments that are
given in the lower-right hand corner of the plot to the interpolated $\log\teff$
values of the variables, the observed values of $\langle P_{ab}\rangle$ and
$\langle P_c\rangle$, which are specified in the top left-hand corner, are
reproduced to three decimal places.  (See Paper I for some discussion of the
rationale behind the introduction of the $\delta\teff$\ parameter.)  That is,
we obtain $\langle\Delta P_{ab}\rangle = 0.000 \pm 0.026$~d and $\langle\Delta
P_c\rangle = 0.000 \pm 0.025$~d.
 
If all of the outliers that are identified in Figs.~\ref{fig:hb085}
and~\ref{fig:per085} had been removed from the sample, the resultant values of
$\langle\Delta P_{ab}\rangle$ and $\langle\Delta P_c\rangle$\ would have been
$0.002 \pm 0.012$~d and $0.000 \pm 0.013$~d, respectively (without applying
any adjustment to the derived temperatures).  Clearly, there is very good
consistency between the predicted and observed periods for the majority of the
RR Lyrae in NGC\,6362 if the cluster has [Fe/H] $\approx -0.85$.  In fact, this
was quite an unexpected result because \citet{smk17} have reported that the
light curves of 69\% of the RRab stars and 19\% of the RRc stars exhibit the
Blazhko effect.  It would seem that this is not a serious complication for most
of these stars though this may provide at least a partial explanation of the
seemingly anomalous periods (or CMD locations) of some of the outliers in
Figs.~\ref{fig:hb085} and \ref{fig:per085}, such as V12 and V13.   
 

Before considering a lower metallicity, some additional remarks concerning
Fig.~\ref{fig:hb085} are warranted.  In particular, the fairly sharp boundary
between the fundamental and first-overtone pulsators seems contrary to
expectations if the hysteresis effect (\citealt{vab73}) is a real phenomenon.
If there was any significant delay in the transformation of an RRab star into
an RRc star during the evolution from red to blue, and vice versa, the boundary
between the $ab$- and $c$-type variables should be bluer at fainter values of
$M_V$ than at higher luminosities, resulting in some overlap of the colors of
the fundamental and first overtone pulsators.  However, Fig.~\ref{fig:hb085}
gives the impression that the transition in the pulsation mode occurs at very
nearly the same color regardless of the direction of evolution inside the
instability strip; i.e., there does not appear to be a hysteresis effect.
(This issue has been discussed much more thoroughly in connection with
the RR Lyrae in M\,5 by \citealt[see their section 4.2]{alb16}.)

\begin{figure}[t]
\plotone{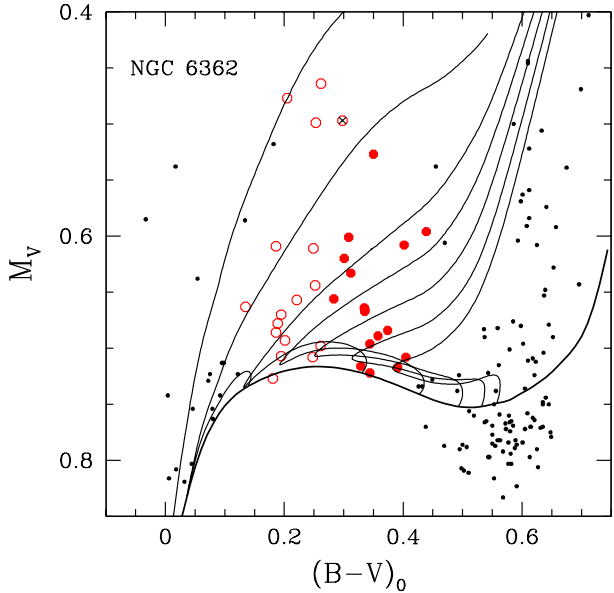}
\caption{Similar to Fig.~\ref{fig:hb085}, except that $(m-M)_V = 14.61$ and
$E(B-V) = 0.070$ have been assumed, and the observations are compared with
stellar models for [Fe/H] $= -1.00$.  The HB tracks, in the direction from left
to right, are for $0.580 \msol$, $0.590 \msol$, and $0.600 \le {\cal M}/\msol
\le 0.620$, in $0.005 \msol$\ increments.  As explained in the text, ZAHB models
for $Y \approx 0.273$ would provide a better match to the faintest RR Lyrae than
those for $Y=0.270$.  To provide a reasonable approximation to the former, the
ZAHB and HB tracks for $Y=0.270$ have been adjusted by $\delta M_{\rm bol} =
-0.008$\ mag.}
\label{fig:hb100}
\end{figure}

Turning to the possibility that NGC\,6362 has [Fe/H] $\approx -1.0$: a magnified
version of the middle panel of Fig.~\ref{fig:hb3} indicates that we would need
to fit ZAHB models for a helium abundance slightly greater than $Y=0.270$ to the
observations in order to match the faintest RR Lyrae.  Rather than compute a
new grid of models for the optimum helium abundance (estimated to be $Y =
0.273$), we opted to shift the $Y=0.27$ grid for [Fe/H] $= -1.0$ by
$\delta M_{bol} = -0.008$\ mag so as to achieve the fit to the data that is
shown in Fig.~\ref{fig:hb100}.   
The main consequence of this approximation is that the inferred masses of the
variable stars will be too small by a few thousandths of a solar mass, but this
will introduce only a small error ($\lta 0.003$~d) in the predicted periods.
Larger errors will certainly arise from the assumption of constant $Y$, as it
seems likely that some of the variables will have appreciably higher helium
abundances.  Still, the increased distance modulus that is associated with a
reduction in [Fe/H] from $-0.85$ to $-1.0$ will have a much larger effect on the
predicted periods than those arising from star-to-star helium abundance
variations.
%

In fact, increasing the adopted value of $(m-M)_V$ by 0.05 mag results in larger
periods by $\sim 0.02$~d, while the net effect on the periods of changes to the 
value of $Z$ and to the interpolated temperatures and masses is very much
smaller.  Thus, if the properties of the variables are obtained by interpolating
in the models for [Fe/H] $= -1.0$ ($Z = 2.626\times 10^{-3}$), we obtain 
$\langle\Delta P_{ab}\rangle = 0.018 \pm 0.029$~d and $\langle\Delta P_c\rangle
= 0.016 \pm 0.027$~d for the mean differences between the predicted and observed
periods of the 17 RRab and 16 RRc stars in our sample.  It would be possible to
reduce these offsets to 0.0 by, e.g., making an adjustment to the temperatures
of the variables amounting to $\delta\teff = 0.0040$ for the RRab pulsators and
$\delta\teff = 0.0065$ for the RRc stars --- offsets that are within the
$1\,\sigma$ uncertainties of the model $\teff$\ scale.  

Alternatively, an increased reddening by only 0.01 mag can accomplish almost the
same thing.  That is, if we were to adopt $E(B-V) = 0.08$ for NGC\,6362 and then
compute the periods of its RR Lyrae from their interpolated properties, we would
obtain $\langle\Delta P_{ab}\rangle = 0.001\pm 0.028$ and
$\langle\Delta P_c\rangle = 0.009 \pm 0.025$ days.  Thus, even though the
apparent distance modulus is virtually independent of $E(B-V)$, because it is
based on a fit of ZAHB models to the reddest HB stars where the ZAHB is nearly
horizontal, there is sufficient uncertainty in the reddening that, with a small
adjustment to the adopted $E(B-V)$ value, it is possible to obtain satisfactory
agreement between the predicted and observed periods over a fairly wide range
in [Fe/H] (from, say, $\sim -0.8$ to $\sim -1.1$).  Whether or not simulations
of the entire HB population are able to place tighter constraints on the cluster
metallicity is examined in the next section.

\begin{table}[b]
\begin{center}
\caption{Fitted Parameters of NGC\,6362 from Simulations of its HB}
\label{tab:fitn6362}
\begin{tabular}{cccccc}
\hline
\noalign{\vskip 2pt}
\hline
\noalign{\vskip 1pt}
 Population ($i$) &  $f_i$ & $Y_i$ & ${\cal M}_i/{\cal M}_\odot$ &
 $\Delta {\cal M}_i/{\cal M}_\odot$ & $\sigma_i/{\cal M}_\odot$ \\ [0.5ex]
\hline
\multicolumn{6}{c}{[Fe/H] $= -0.85$} \\ [0.5ex]
 $1$ & $0.36$ & $0.250$ & $0.860$ & $0.240$ & $0.01$ \\
 $2$ & $0.34$ & $0.265$ & $0.835$ & $0.245$ & $0.01$ \\
 $3$ & $0.30$ & $0.280$ & $0.815$ & $0.243$ & $0.01$ \\
\noalign{\vskip 1pt}
\multicolumn{6}{c}{[Fe/H] $= -1.00$} \\ [0.5ex]
 $1$ & $0.45$ & $0.250$ & $0.837$ & $0.195$ & $0.01$ \\
 $2$ & $0.45$ & $0.265$ & $0.813$ & $0.212$ & $0.01$ \\
 $3$ & $0.10$ & $0.280$ & $0.790$ & $0.208$ & $0.01$ \\
\hline
\end{tabular}
\end{center}
\end{table}

\subsection{Simulations of the NGC\,6362 HB}
\label{subsec:n6362hbsim}

\begin{figure*}[t]
\plotone{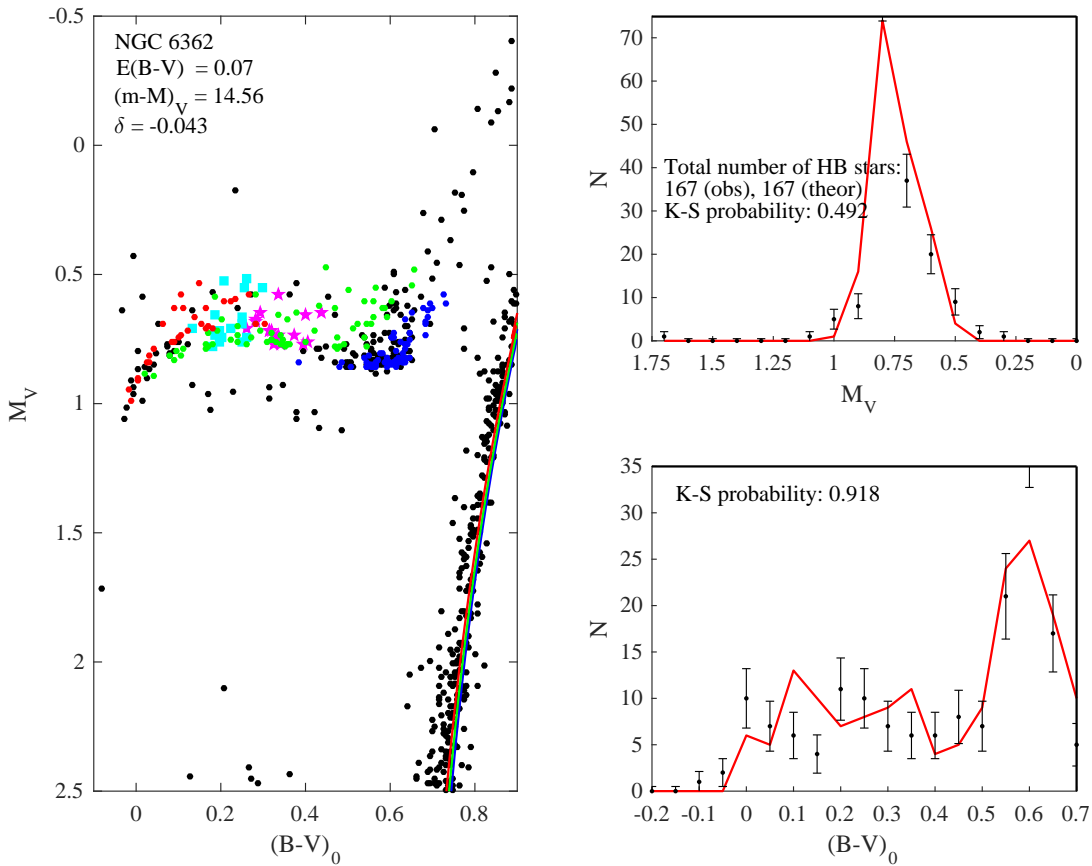}
\caption{As in Fig.~\ref{fig:syn6809}, except that a synthetic HB for NGC\,6362
has been generated from tracks for [Fe/H]\,$=-0.85$, [$\alpha$/Fe]\,$=0.4$, and
$Y = 0.25$, 0.265, and 0.28 (blue, green, and red filled circles, respectively.
Superimposed on the cluster giants are evolutionary tracks for 0.86, 0.835,
and $0.815 \msol$ for the same three values of $Y$, in turn; see the discussion
of similar models in connection with Fig.~\ref{fig:syn6809}.}
\label{fig:syn6362}
\end{figure*}

Using the same procedures that were briefly summarized in
\S\,\ref{subsec:m55hbsim}, we have generated synthetic HB populations applicable
to NGC\,6362 from evolutionary tracks for [Fe/H] $= -0.85$ and $-1.0$ assuming,
in both cases, [$\alpha$/Fe] $= +0.4$ and $Y = 0.25$, 0.265, and 0.28.  (For
these simulations, the photometric uncertainty was taken to be
$\sigma_\mathrm{phot}\approx 0.005$ mag, as determined for the brightest stars
in  the sample of nearly 10,000 proper-motion selected members that were observed
by \citealt{zkr12}.\footnote{http://case.camk.edu.pl/results/ProperMotions/}) 
With the fitting parameters set to the values listed in Table~\ref{tab:fitn6362},
the simulated HBs for metallicities that differ by 0.15 dex provide equally
satisfactory fits to the observations, according to the calculated
Kolmogorov-Smirnov probability.  For instance, the best fit of our synthetic HB
for [Fe/H] $= -0.85$\ to the observed distribution of RR Lyrae stars from
\citet{okt01} and non-variable stars from Zloczewski et al.~is shown in
Figure~\ref{fig:syn6362}.  If the simulations are based instead on the models
for [Fe/H] $= -1.0$, the calculated K-S statistic for our best matches to the
observed numbers of stars as functions of absolute magnitude and $(B-V)_0$
color are, in turn, 0.612 and 0.791 (i.e., just slightly higher and lower,
respectively, than the values reported in Fig.~\ref{fig:syn6362}). 

However, in order to achieve this, stars with [Fe/H] $=-0.85$ would have to 
lose more mass along the RGB and a higher fraction of the stars in NGC\,6362
would need to belong to the most helium-rich population (see
Table~\ref{tab:fitn6362}).  This is a possible problem, as a mean mass loss of
$\sim 0.24\,\msol$\ is significantly higher than we have found for both more
metal-poor and more metal-rich GCs that we have studied so far in this series
of papers.  Indeed, the predicted mass loss if [Fe/H] $= -1.0$ agrees very well
with the estimate of $\Delta{\cal M}_i\approx 0.21\,{\cal M}_\odot$ that is
obtained for a star with an initial mass ${\cal M}_i\approx 0.825\,{\cal
M}_\odot$ and the adopted chemical abundances if $\eta_\mathrm{R} = 0.45$ is
assumed in Reimers' mass-loss formula. (Recall our demonstration in Paper I that
differences in $Y_i$ should not affect the Reimers mass-loss estimate.)  This
suggests that the metallicity of NGC\,6362 {\it may} be closer to $-1.0$ than
to $-0.85$.

The fits of theoretical distributions of HB stars to the observations are quite
sensitive to variations in the fractions of their multiple populations $f_i$.
For both NGC\,6362 and M\,55, a 5--10\% change in the fractions of the first two
populations, those for $Y_1 = 0.25$ and $Y_2 = 0.265$, results in a significant
reduction of the K-S probabilities.  In the case of NGC\,6362, it is the K-S
probability for the color fit that is the most sensitive to variations of $f_1$
at a fixed value of $f_3$.  For instance, although our best HB fit on the 
assumption of [Fe/H] $= -1.0$ yielded $f_1 = 0.45$ and $f_3 = 0.10$ (see 
Table~\ref{tab:fitn6362}), comparatively good fits are also obtained for $f_1 =
0.50$ at both $f_3 = 0.10$ and $f_3 = 0.05$.  However, larger variations of
$f_1$ lead to significant reductions in the K-S probability for the color fit
(by more than a factor of two, and increasing with the variation).  A similar
result is obtained if the adopted metallicity is [Fe/H] $= -0.85$.  The HB fits
are much more sensitive to variations in the RGB mass loss, for which changes
of a few percent lead to significant reductions of the K-S probabilities. 

In excellent agreement with the distance moduli that are obtained from fits of
ZAHB models to the observed HB (see Figs.~\ref{fig:hb3}, \ref{fig:hb085}, and
\ref{fig:hb100}), our simulations yield $(m-M)_V = 14.56$ if [Fe/H] $=-0.85$ and
$(m-M)_V = 14.60$ if [Fe/H] $= -1.0$.  For both of these determinations, the
adopted reddening is $E(B-V) = 0.07$.  By comparison, in a paper that was
submitted for publication concurrently with ours, \citet{aab18} derived $E(B-V)
= 0.063 \pm 0.024$, [Fe/H] $ = -1.066 \pm 0.126$, and $(m-M)_V = 14.69 \pm 0.08$
from new time-series CCD $VI$ photometry of the $ab$-type RR Lyrae (with very 
similar results for the RRc stars).  To within the uncertainties, these results
are fully consistent with our determinations.

\subsection{The Age of NGC\,6362}
\label{subsec:age6362}

\begin{figure}[t]
\plotone{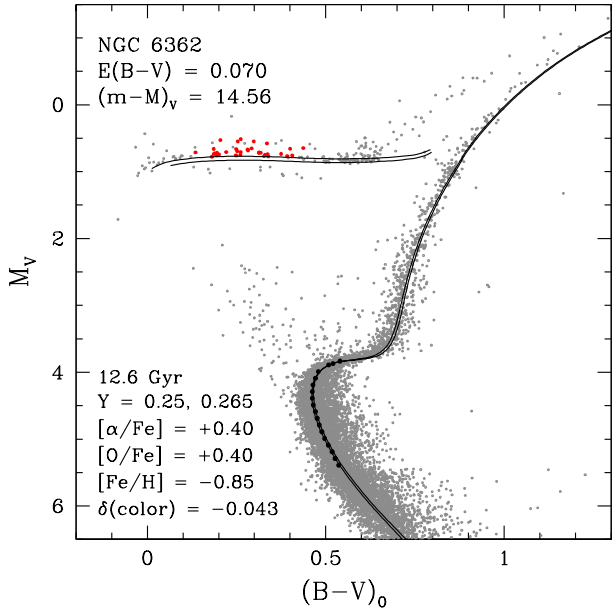}
\caption{Similar to Fig.~\ref{fig:bvm18}, except that ZAHBs and 12.6 Gyr
isochrones for the indicated helium and metal abundances have been fitted to the
CMD of NGC\,6362 (from \citealt{zkr12}), on the assumption of the reddening and
apparent distance modulus that are specified in the top left-hand corner.}
\label{fig:i085}
\end{figure}

Since the binaries in NGC\,6362 are comprised of stars that are located near
the cluster turnoff, we need to identify which isochrones should be compared
with their properties.  The necessary next step in our analysis is therefore the
determination of the age of NGC\,6362 on the assumption of our best estimates
of its reddening and distance modulus.  As in the case of M\,55 (see
\S\,\ref{sec:m55}), the median fiducial sequence for stars in the vicinity of
the turnoff was derived in the usual way (see Paper II) so that the procedure
used to select the best-fit isochrone involves very little, if any, subjective
errors.

Figure~\ref{fig:i085} shows that 12.6 Gyr isochrones for [Fe/H] $= -0.85$, 
[$\alpha$/Fe] $= +0.4$, and $Y = 0.25$, 0.265 provide very good fits to the
turnoff of NGC\,6362 if $E(B-V) = 0.070$ and $(m-M)_V = 14.56$ (as derived in
the preceding sections).  The relatively small change in $Y$ clearly has a much
bigger impact on models for core He-burning stars than on those for the MS and
RGB phases of evolution. Although Fig.~\ref{fig:i085} gives the impression that
the observed subgiant branch is somewhat more steeply sloped than those of the
isochrones, this may be the result of whatever is causing the predicted RGB to
be offset slightly to the red of the cluster giants.  For instance, if we forced
the predicted giant branch to match the observed one by, among other
possibilities, allowing for a small increase in the mixing-length parameter
between the TO and the RGB or making suitable adjustments to the atmospheric
boundary condition, the SGB slope discrepancy would no longer be apparent.  

\begin{figure}[t]
\plotone{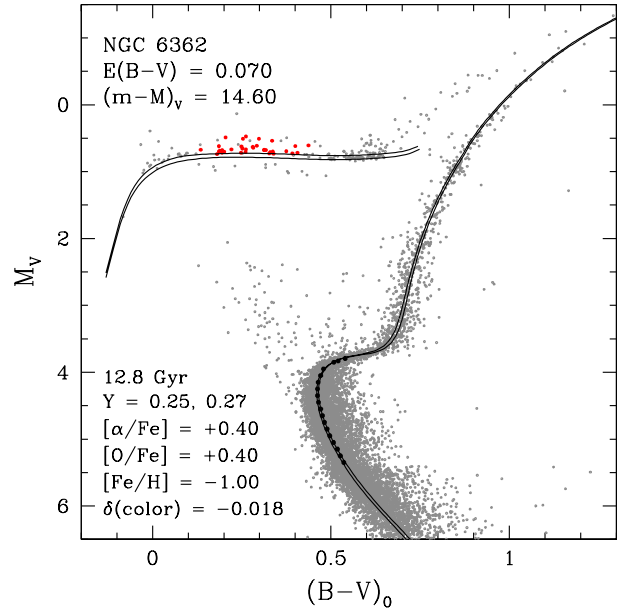}
\caption{As in the previous figure, except that models for [Fe/H] $= -1.0$ are
compared with the observations.}
\label{fig:i10}
\end{figure}

On the other hand, a lower metallicity can accomplish the same thing.  As shown
in Figure~\ref{fig:i10}, 12.8 Gyr isochrones for [Fe/H] $= -1.0$ provide a
significantly improved fit to the cluster CMD.  If a small zero-point adjustment
is applied to the isochrone colors ($-0.018$ mag), the models reproduce the
locations and the slopes of the MS, SGB, and RGB populations rather
well.\footnote{The color offset could be the result of errors in, e.g., the
adopted color--$\teff$ relations, the model $\teff$\ scale, or the assumed
cluster properties.  Generally, they amount to $\lta 0.02$ mag (see Papers I and
II, as well as VBLC13), so it is a concern that the fitting of isochrones for
[Fe/H] $= -0.85$ to the CMD of NGC\,6362 (see Fig.~\ref{fig:i085}) requires
a much larger adjustment (0.043 mag).  This suggests that NGC\,6362 may not
have such a high metallicity.}  Even so, this is not a compelling argument in
support of this possibility because there are so many factors, each with
significant uncertainties, that impact the model $\teff$\ and color scales.
Because turnoff ages depend quite strongly on the total C$+$N$+$O abundance
(\citealt{vbd12}), NGC\,6362 could easily be several hundred Myr younger or
older than the ages given in Figs.~\ref{fig:i085} and~\ref{fig:i10}.  For
instance, if the cluster stars have [O/Fe] $= +0.6$, to be consistent with
recent findings for field stars of similar metallicity (e.g., \citealt{zmy16}),
the turnoff age would be $\approx 0.5$ Gyr less than that derived from models
for [O/Fe] $= +0.4$ (assuming the same distance modulus and [$\alpha$/Fe]
$= 0.4$ for the abundances of the other $\alpha$-elements).  However, because
different C$+$N$+$O abundances will affect the mass-luminosity relation of the
best-fit isochrone, we should be able to use the eclipsing binaries in NGC\,6362
to discriminate between the possible metal abundance mixtures (at least in
principle).

\subsection{The Binaries V40 and V41 in NGC\,6362}
\label{subsec:bin6362}

\begin{table}[b]
\begin{center}
\caption{Basic Properties of V40 and V41}
\label{tab:t5}
\begin{tabular}{lcc}
\hline
\hline
\noalign{\vskip 1pt}
Property & Primary & Secondary \\
\hline
\noalign{\vskip 1pt}
\multicolumn{3}{c}{\bf V40} \\ [0.5ex]
 Mass ($\cal M_\odot$) & $0.8337 \pm 0.0063$ & $0.7947 \pm 0.0048$ \\
 Radius ($R_\odot$)    & $1.3253 \pm 0.0077$ &
                         $0.997\phantom{0} \pm 0.013\phantom{0}$ \\
 $V$ magnitude         & $18.698 \pm 0.020\phantom{0}$ &
                         $19.338 \pm 0.027\phantom{0}$ \\ 
 $B-V$ color           & $0.542\phantom{0} \pm 0.023\phantom{0}$ &
                         $0.556\phantom{0} \pm 0.034\phantom{0}$ \\
\noalign{\vskip 1pt}
\multicolumn{3}{c}{\bf V41} \\ [0.5ex]
 Mass ($\cal M_\odot$) & $0.8215 \pm 0.0058$ & $0.7280 \pm 0.0047$ \\
 Radius ($R_\odot$)     & $1.0739 \pm 0.0048$ & $0.7307 \pm 0.0046$ \\
 $V$ magnitude         & $19.089 \pm 0.017\phantom{0}$ &
                         $20.274 \pm 0.018\phantom{0}$ \\
 $B-V$ color           & $0.550\phantom{0} \pm 0.018\phantom{0}$ &
                         $0.650\phantom{0} \pm 0.022\phantom{0}$ \\
\noalign{\vskip 1pt}
\hline
\end{tabular}
\end{center}
\end{table}

\begin{figure*}[t]
\plotone{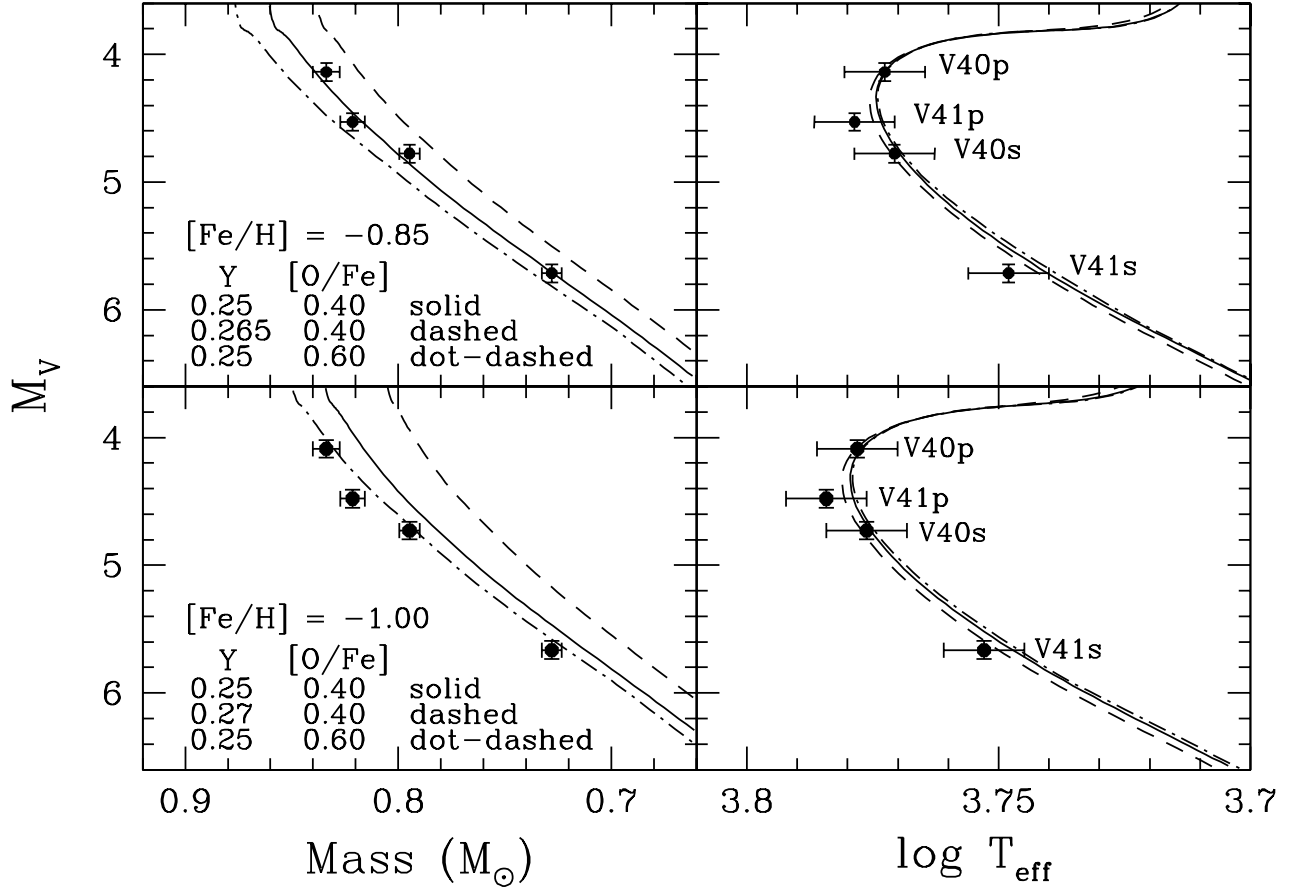}
\caption{{\it Left-hand panels}: superposition of the mass--$M_V$ relations
predicted by the isochrones that were fitted to the CMD of NGC\,6362 in
Figs.~\ref{fig:i085} and \ref{fig:i10} onto the derived masses and absolute $V$ 
magnitudes of the components of V40 and V41.  The dot-dashed isochrones, which
assumes [O/Fe] $= 0.6$ and 0.4 dex enhancements of the other $\alpha$-elements,
were computed for ages of 12.1 and 12.3 Gyr for the [Fe/H] $ = -0.85$ and $-1.0$
cases, respectively.  {\it Right-hand panels}: as in the left-hand panels,
except that the comparisons are made on the $\log\teff$--$M_V$ plane.}
\label{fig:bin4}
\end{figure*}

\citet{ktd15} determined the main properties of the detached, eclipsing binaries
V40 and V41 in NGC\,6362; their results for those quantities that are used in
this investigation are listed in Table~\ref{tab:t5}.  Note that the masses of
the binary components are known to within 0.76\%, whereas the derived radii have
uncertainties amounting to $< 1.3$\%.  Using the same procedures that are
described in detail in \S\,\ref{subsec:v54}, we can evaluate the luminosities of
the components of V40 and V41, and then calculate their temperatures from the
resultant luminosities and the tabulated radii.  The same isochrones that were
discussed in the previous section are compared with the binaries on the
mass--$M_V$\ and $\log\,\teff$--$M_V$\ diagrams in Figure~\ref{fig:bin4}.

Unfortunately, as already shown in our study of M\,55 (see Figs.~\ref{fig:mmv}
and~\ref{fig:mmvmod}), mass--$M_V$ relations are much more dependent on the
helium abundance than on [O/Fe].  An increase in $Y$ by only $\sim +0.0075$
(half of the separation between the solid and dashed loci in the upper 
left-hand panel of Fig.~\ref{fig:bin4}) causes a slightly larger increase in
the mass at a given absolute magnitude than a $+0.2$\ dex increase in [O/Fe]
(the separation between the solid and dot-dashed curves).  Since the helium
contents of stars in NGC\,6362 must vary by at least $\delta Y = 0.015$--0.03
in order to explain the observed HB (in good agreement with the helium
abundance variation derived by \citealt{gcb10}), V40 and V41 could be members
of the population with $Y \approx 0.250$ or those with $Y \approx 0.265$--0.28
or any intermediate helium abundance.  Taken at face value, the upper left-hand
panel suggests that the binaries have $Y=0.25$ and [$\alpha$/Fe] $= 0.4$ if they
have [Fe/H] $\approx -0.85$, whereas the lower left-hand panel indicates a
preference for the dot-dashed isochrone among those computed for [Fe/H]
$= -1.0$; this also assumes $Y=0.250$, but [O/Fe] $= 0.6$. 

The difficulty is that the observations could be explained equally well by 
different combinations of $Y$ and [$\alpha$/Fe].  For example, an isochrone for
$Y=0.26$, [O/Fe] $= 0.6$, and [Fe/H] $= -0.85$ would fit the data just as well
as the solid curve in the upper left-hand panel.  Nevertheless, it does appear
that a metallicity less than [Fe/H] $= -1.0$, such as the latest estimate
of $-1.07$ from high-resolution spectroscopy (\citealt{mmd17}), 
would be difficult to accommodate
because that would require $Y < 0.250$, which is unlikely given that $Y=0.250$
is very close to current best estimates of the primordial helium abundance
(e.g., \citealt{cfo16}), or [O/Fe] $> 0.6$, which also seems improbable as
spectroscopic studies of stars with $-1.2 \lta$ [Fe/H] $\lta -0.8$ generally
find [O/Fe] $= 0.6$ or less (e.g., \citealt{rmc12}, \citealt{zmy16}).  Indeed,
low $\alpha$-element abundances (i.e., [$\alpha$/Fe] or [O/Fe] $\lta 0.2$) would
also be problematic for any [Fe/H] $\le -0.85$ and $Y \gta 0.25$.

The right-hand panels of Fig.~\ref{fig:bin4} show that the temperatures of the
binary components, as calculated from the luminosities implied by the adopted
distance moduli and the observed radii, are in rather good agreement with the
model $\teff$\ scale, just as we found in the case of M\,55.  In fact, slightly
higher temperatures are favored, though not as high as those given by the
IRFM-based $(B-V)_0$--$\teff$--[Fe/H] relation of CRMBA.  Assuming
$E(B-V) = 0.07$ and the $B-V$ colors that are listed in Table~\ref{tab:t5}, that
relation yields higher effective temperatures by $\delta\log\teff = 0.010$--0.018
for the four binary components if they have [Fe/H] $= -0.85$, or $\delta\log\teff =
0.008$--0.016 assuming [Fe/H] $= -1.0$.  This is not a serious concern, however,
since the $1\,\sigma$ error bars of these independent determinations overlap,
though just barely.  We note that errors in the observed colors, which have
relatively high uncertainties (see Table~\ref{tab:t5}), could explain about
one-half of the discrepancies. 

It is important to appreciate that any increase in the ZAHB-based values of
$(m-M)_V$\ would imply higher luminosities for V40 and V41 and therefore higher
effective temperatures.  Moreover, since our models assume $Y=0.250$, and because
they take the gravitational settling of helium into account, which results in
reduced He abundances at the RGB tip than the predictions of non-diffusive
models, resulting in fainter ZAHB models, the apparent distance moduli derived
in this study will be close to their minimum possible values.  As a result, the
$\teff$ values that we determined for the binary components must be close to
their minimum possible values as well.  

\begin{figure}[t]
\plotone{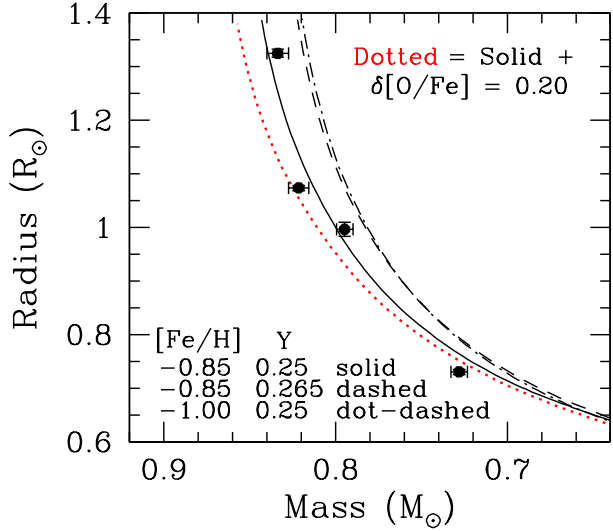}
\caption{Similar to the previous figure, except that isochrones for the 
indicated chemical abundances are compared with the observed masses and radii
of V40 and V41.}
\label{fig:mrm085}
\end{figure}

As expected, comparisons of several of the same isochrones with the observations
on the mass--radius diagram (see Figure~\ref{fig:mrm085}) look very similar to
those shown in Fig.~\ref{fig:bin4}, thereby reinforcing the conclusions 
discussed above.  Interestingly, both figures give the impression that V40 and
V41 follow somewhat different ${\cal M}$--$R$ relations, even on the
$(\log\teff,\,M_V)$-plane despite the large error bars attached to the derived
temperatures.  \citet{ktd15} noticed the same thing, but whereas they tentatively
suggested that this was due to a $\sim 1.5$~Gyr age difference, the more likely
explanation is that V40 has a slightly lower initial helium abundance than V41
(by only $\Delta Y \sim 0.007$) given that the cluster HB shows unambiguous
evidence for a much larger spread in $Y$.

Based on the results presented in Figs.~\ref{fig:bin4} and \ref{fig:mrm085},
the binaries in NGC\,6362 suggest that the cluster has [Fe/H] $= -0.90 \pm 0.10$
and an oxygen abundance that is somewhere in the range $+0.4 \le$ [O/Fe] $\le
+0.6$.  The effect on the predicted age of adopting a lower metallicity by 0.05
dex and a higher value of [O/Fe] by 0.1, relative to the metal abundances
that were assumed in Fig.~\ref{fig:i085}, is a reduction by about 0.2 Gyr.  That
is, our best estimate of the age of NGC\,6362 is 12.4 Gyr.  As for M\,55, the
uncertainty of this determination is $\approx \pm 0.8$\ Gyr, of which $\pm
0.5$\ Gyr is due to a $\pm 0.05$\ mag uncertainty in the distance modulus and
the rest corresponds to the net effect of metal abundance uncertainties.

Our results for M\,55, NGC\,6362, and the GCs considered in Papers I and II
clearly depend on the reliability of our stellar models, especially 
those for the HB phase since they provide the basis for our adopted distance
moduli.  Consequently, it is important to check how well the predicted $(m-M)_V$
values agree with determinations based on other considerations.  This is the
subject of the next section.

\section{Distance Constraints from Standard Candles}
\label{sec:candle} 

There have been a number of recent developments concerning the use of RR Lyrae
and solar neighborhood subdwarfs as standard candles, but their
implications for GC distances have not been very encouraging.  In particular,
the trigonometic parallaxes that have been derived for field RR Lyrae from
first-epoch {\it Gaia} observations (\citealt{cer17}) seem to favor a much
shorter distance scale than the latest fits of GC main sequences to nearby
subdwarfs; see \citet{cmo17} and \citet{ogc17}, who made use of both the
parallaxes that they derived from data taken with the {\it HST} Fine Guidance
Sensors and those obtained from the {\it Gaia} mission (\citealt{bvp16}) for a
small number of stars.  Since our HB models satisfy the RR Lyrae constraint
quite well, as reported in Paper I, the main focus of this section will be on
the subdwarf-fitting method.  However, we will first compare the mean absolute
magnitudes that we have derived for the RRab stars in several GCs with various
empirical $M_V$--[Fe/H] relationships.

\subsection{The RR Lyrae Standard Candle}
\label{subsec:rrlc}

\begin{figure}[t]
\plotone{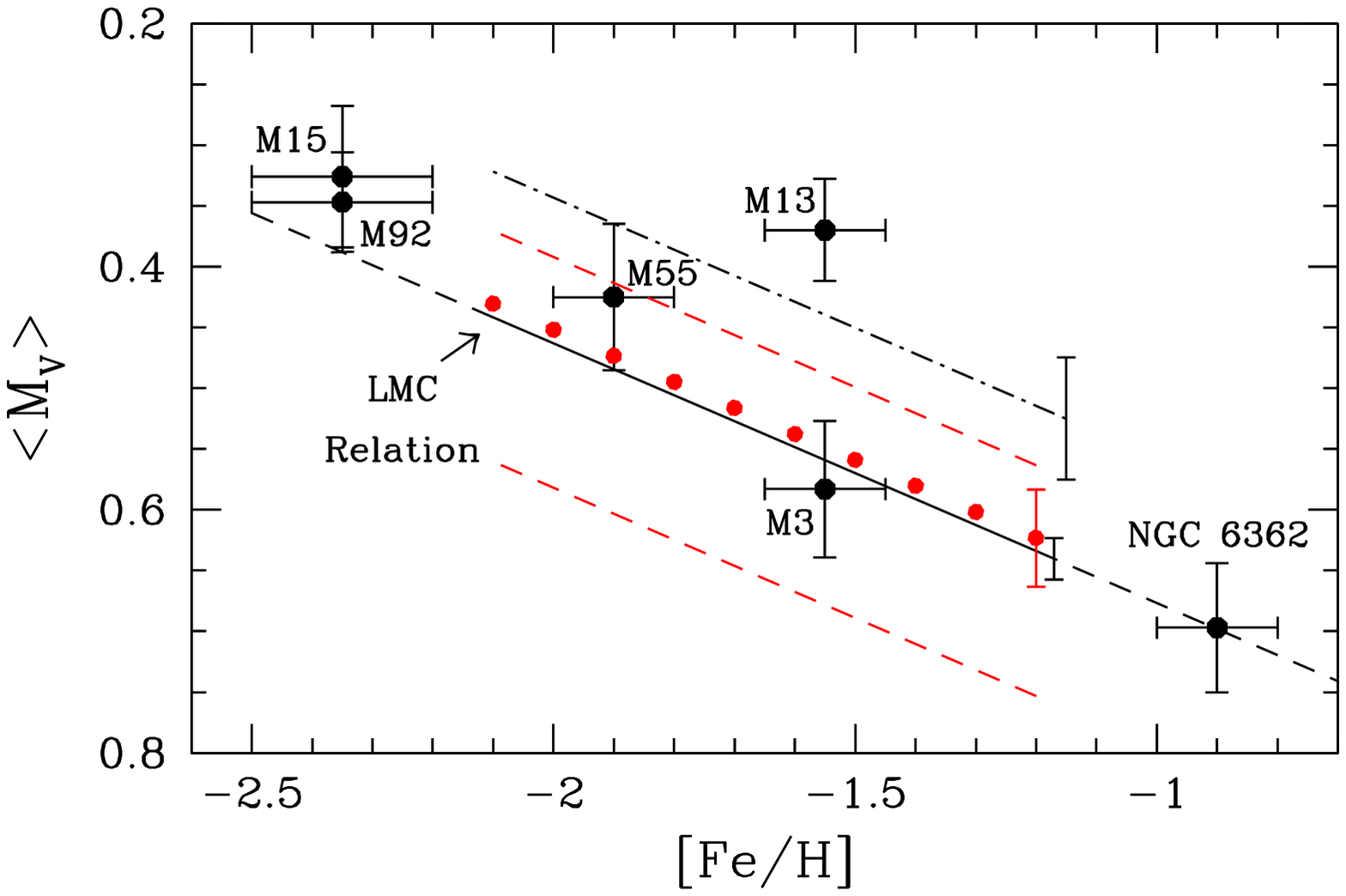}
\caption{RR Lyrae $M_V$ versus [Fe/H] relations from \citet[solid line]{cgb03},
\citet[dot-dashed line]{bmf11}, and \citet[filled circles and dashed lines in
red]{cer17} together with our determinations of $\langle M_V\rangle$ for the
variable stars in 6 GCs (filled circles with error bars).  Black dashed lines
provide linear extrapolations of the solid line to metallicities outside the
range occupied by nearly all of LMC RR Lyrae that were observed by \citet{cgb03}.
These results assume that the true distance modulus of the LMC is $(m-M)_0 =
18.496$ (\citealt{pgg13}).  Note that the investigations by \citet{bmf11} and
\citet{cer17} both assumed $\Delta M_V/$[Fe/H] $= 0.214$, as found by
\citet{cgb03}.}
\label{fig:mvfeh}
\end{figure}

Figure~\ref{fig:mvfeh} plots the mean absolute magnitudes that we have derived
for 6 GCs in Papers I--III of the present series as a function of their adopted
metallicities.  In order for our simulated HBs to match the properties of the
observed ones, including the periods of the cluster RR Lyrae, these $\langle
M_V\rangle$\ values cannot be in error by more than $\sim \pm 0.05$\ mag (the
adopted error bar) {\it if} our models for the core He-burning phase are
trustworthy.  Current [Fe/H] estimates for most clusters are probably accurate
to within $\sim\pm 0.1$\ dex ($1\,\sigma$), though uncertainties closer to
$\sim\pm 0.15$\ dex appear to be more reasonable at the lowest metallicities 
(recall the discussion in \S\ref{sec:intro}).  Accordingly, the horizontal error
bars are larger for M\,15 and M\,92 than for the other clusters that are 
identified in Fig.~\ref{fig:mvfeh}.  Superimposed on these data are several
lines representing empirical calibrations that will now be discussed in turn.

The solid curve, which is defined by the equation $\langle M_V\rangle = 0.57 +
0.214($[Fe/H] $+1.5)$, is widely considered to be the best empirical calibration
of the RR Lyrae standard candle at the present time.  This equation follows from
the observed dependence of $\langle V_0\rangle$ on [Fe/H] that was obtained by
\citet{cgb03} from their observations of $\sim 100$ variables in the Large
Magellanic Cloud (LMC), if the very accurate LMC distance determined by
\citet{pgg13} from eclipsing binary stars is adopted.  (The dashed lines are
simply linear extensions of the solid curve outside the range in [Fe/H] spanned
by the majority of the RR Lyrae in the Clementini et al.~sample.)  Incidently,
our ZAHB models for [Fe/H] $=-2.0$ and $-1.4$, assuming $Y=0.25$, predict nearly
the same slope as the solid curve, specifically, $\Delta M_V/$[Fe/H] $= 0.220$
(versus 0.214; as noted at the beginning of this paragraph).  

Using the {\it Hubble Space Telescope} Fine Guidance Sensors, \citet{bmf11}
determined the trigonometric parallaxes of 5 RR Lyrae in the solar neighborhood,
from which the dot-dashed relation and the attached error bar were derived.  It
assumes the same slope as the solid line, but a brighter zero-point by 0.12 mag.
Perhaps the main concern with these results, aside from the small number
of stars in the sample and the large uncertainties in the derived $M_V$ values
(generally $\gta 0.15$\ mag, except in the case of RR Lyr itself) is the
adoption of selection bias corrections (\citealt{lk73}) ranging from $-0.02$\ mag
to $-0.11$\ mag.  Such corrections should not be applied to these stars
according to \citet{fr14}.  

The possibility that one (or two?) of the brightest stars are in a very
different evolutionary state than the others would also have a significant
effect on the derived $\langle M_V\rangle$--[Fe/H] relation when there are only
a few variables in the sample, as in the \citet{bmf11} study.  This concern is
exemplified by our results for M\,3 and M\,13, which appear to have close to the
same metallicity and age, but the mean absolute magnitudes of their RR Lyrae
populations differ by more than 0.2 mag; see Fig.~\ref{fig:mvfeh}.  This
luminosity difference arises mainly because all of the RR Lyrae in M\,13 are
believed to be highly evolved stars from ZAHB locations well to the blue of the
instability strip whereas a large fraction of the M\,3 variables are located
adjacent to the ZAHB.  A difference in helium abundance (see Paper II) would
also contribute to the luminosity offset.  

\citet{cer17} employed three different approaches in their analysis of 
preliminary {\it Gaia} observations for 200 field RR Lyrae (see their paper for
details), finding zero-points that varied from 0.50 mag to 0.69 mag (at [Fe/H]
$= -1.5$), if they assume the same slope of the $\langle M_V\rangle$ vs.~[Fe/H]
relation as in the aforementioned studies.  Their application of a Bayesian
fitting method yielded the relationship that has been plotted as the red filled
circles in Fig.~\ref{fig:mvfeh}; the other two (represented by the red dashed
lines) are offset by $-0.06$\ mag and by $+0.13$\ mag, respectively, at a given
[Fe/H] value.  The faintest of the three relations is considered by Clementini
et al.~to be the least trustworthy one because it involves the direct
transformation of parallaxes into absolute magnitudes, which causes obviously
asymmetric errors in the $M_V$ values when errors are large, besides being
unable to take negative parallaxes into account.  Although their preferred
results agree quite well with those represented by the solid curve, more
definitive findings will, as stated by the authors, have to await further studies
of the systematic errors in the parallax determinations and improvements to the
data that will accompany future releases of {\it Gaia} observations.

As reported in Paper I, and shown in Fig.~\ref{fig:mvfeh}, the mean absolute
magnitude of M\,3 RR Lyrae, assuming the distance modulus predicted by our
ZAHB models and HB simulations, agrees with the value implied by the empirical
$\langle M_V\rangle$--[Fe/H] relation from \citet{cgb03} to within $\sim 0.03$
mag.  It is to be expected that the variables in M\,55, and especially those
in M\,13, would lie above the same relation because they are predicted to be
significantly more evolved stars and, in the case of M\,13, to have somewhat
higher $Y$, than those residing in M\,3 (see Paper II).  Since M\,15 is
especially rich in RR Lyrae, its location somewhat above the ``LMC relation"
(by only $1\,\sigma$, however) could also arise if it has a higher helium 
abundance in the mean than M\,3 (see Paper I) or it may be a consequence of
the assumed metallicity.
If we had adopted [Fe/H] $= -2.5$ for M\,15 (and
M\,92), as found in a number of recent studies (e.g., \citealt{sks11},
\citealt{rs11}), we would have found that the RR Lyrae in both clusters lie on
essentially the same $\langle M_V\rangle$ vs.~[Fe/H] relation as those in M\,3
(see Fig.~\ref{fig:mvfeh}).  Indeed, we intend to examine in a forthcoming study
whether it is possible to obtain a satisfactory explanation of the properties
of the RR Lyrae in M\,15 if they have a normal helium abundance but a very low
metallicity.  (Regardless of what the planned investigation reveals, we believe
that a large spread in $Y$ is needed to explain the extended blue HB tails in
such clusters as M\,15 and M\,13; see Paper II.)

Finally, the variables in NGC\,6362 are predicted to lie just above the
$\langle M_V\rangle$--[Fe/H] relationship that passes through the point
representing M\,3.  This is consistent to well within the respective error bars
with the prediction from our HB simulations that the RR Lyrae in NGC\,6362 have
a higher helium abundance than those in M\,3 by $\Delta Y \sim 0.01$.  In fact,
it is unlikely that the variation of $\langle M_V\rangle$ with metallicity is
strictly linear over the full range in [Fe/H] (see, e.g., the discussion by
\citealt{cat09}); i.e., more metal-rich variables with the same $Y$ and 
evolutionary state are probably fainter than one would infer from a linear fit
to the more metal-poor variables.  Such a quadratic variation would tend to
improve the consistency between theory and the observations of M\,3 and NGC\,6362.
In any case, we can conclude from Fig.~\ref{fig:mvfeh} that our models satisfy
the RR Lyrae constraint very well.  As a result, the distance moduli that we
have derived for GCs in this series of papers, as well as those determined in
the investigation by \citet{vbl13}, should be accurate to within $\sim\pm 0.05$
mag ($1\,\sigma$).

\subsection{The Local Subdwarf Standard Candle}
\label{subsec:sbdc}

\citet{vcs10} have already shown that the IRFM $\teff$\ scale that was derived
by CRMBA for solar neighborhood stars is nearly identical with that predicted
by stellar models.  Moreover, the same study (also see \citealt{bsv10})
demonstrated that the colors of nearby dwarf stars are generally quite well
reproduced when the models are transposed from the 
$(\log\,\teff,\,M_V)$-diagram to various CMDs using color transformations based
on MARCS model atmospheres and synthetic spectra (specifically, the
color--$\teff$\ relations presented in the subsequent study by \citealt{cv14}).
This indicates that the predicted photometric properties of MARCS model
atmospheres (\citealt{gee08}) are able to match those of local dwarfs only if
the temperatures of the latter are close to the values given by CRMBA (given the
similarity of the stellar evolution and IRFM $\teff$\ scales).  The same can be
said of dwarf stars in open and globular clusters since Victoria-Regina
isochrones would not be able to reproduce the CMD morphologies of their MS
populations so well (from their turnoffs at $M_V \sim 4$ to $M_V \gta 8.5$; see
\citealt[their Figs.~9, 14, and 15]{vbf14}) if they predicted much hotter or 
much cooler temperatures.  

\begin{figure}[t]
\plotone{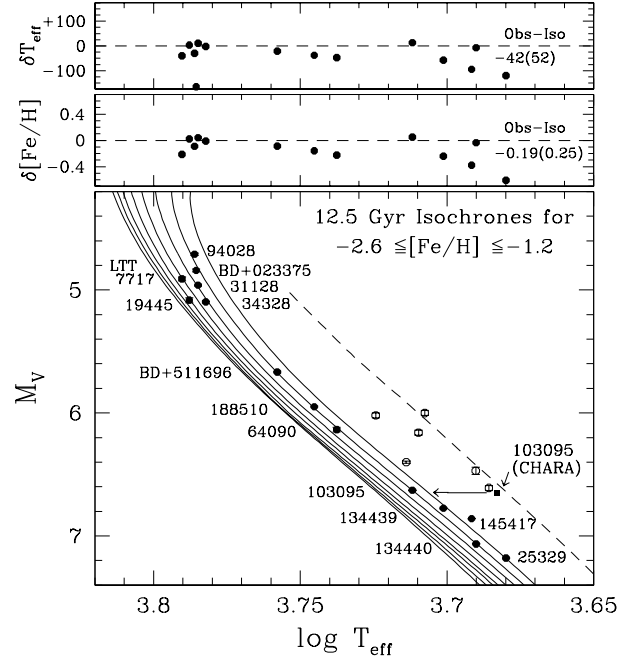}
\caption{{\it Lower panel}: superposition of the properties of nearby subdwarfs
(filled circles) with [Fe/H] $\le -1.3$ and $\sigma(M_V) \le 0.08$\ mag onto
12.5 Gyr isochrones for $-2.6 \le$ [Fe/H] $\le -1.2$, in $+0.2$ dex increments
(from left to right).  Aside from 3 stars with ``BD" or ``LTT" designations,
the subdwarfs are identified by their HD numbers.  Their absolute magnitudes
are based on {Gaia} DR2 parallaxes, except in the case of HD\,19445 (see the 
text) while their metallicities and effective temperatures have been taken from
the study by CRMBA.  The dashed curve represents the lower-MS portion of a solar
abundance isochrone that provides an excellent fit to the morphology of the
$[(V-K_S)_0,\,M_V]$ CMD of M\,67 (\citealt{vbf14}).  The open circles and
vertical error bars, which are comparable to the diameters of the open circles,
indicate the locations of the subdwarfs that were studied by \citet{cmo17},
assuming the properties derived by them.  The filled square gives the location
of HD\,103095 if the temperature derived by \citet{ctb12} from CHARA
interferometric observations is adopted.  {\it Middle panel}: the difference,
for each of the stars that has been plotted as a filled circle, between the
[Fe/H] value adopted by CRMBA and that inferred from the interpolated (or
extrapolated) isochrone that matches its location on
the $(\log\teff,\,M_V)$-diagram (lower panel).  The mean $\delta$\,[Fe/H] value
and the standard deviation are $-0.19$\ dex and 0.25 dex, respectively (as
indicated).  {\it Upper panel}: similar to the middle panel, except that the
difference between the CRMBA estimates of $\teff$\ of each star and that
predicted by the isochrone for its metallicity at the observed $M_V$\ value.}
\label{fig:sbdte}
\end{figure}

It is worthwhile to revisit these findings in view of the very significant
improvements that have been made to the absolute magnitudes of many of the most
metal-deficient subdwarfs with the recent release of {\it Gaia} DR2
parallaxes.  On the assumption of the effective temperatures given by
CRMBA, the lower panel of Figure~\ref{fig:sbdte} compares the locations on the
$(\log\teff,\,M_V)$-plane of 14 of the most metal-deficient subdwarfs (the
filled circles) with Victoria-Regina isochrones (\citealt{vbf14}) for $-2.6 \le$
[Fe/H] $\le -1.2$.  Although vertical error bars have been plotted, they
are barely discernible, if at all, which indicates that the $M_V$ values are
now sufficiently well determined that distance uncertainties are
inconsequential; i.e., comparisons between theory and observations will be much
more dependent on chemical compostion and $\teff$\ uncertainties.  Note that,
because {\it Gaia} DR2 parallaxes are not available for HD\,19445 at this
time, we adopted the DR1 result (\citealt{bvp16}) for this star.  Encouragingly,
all of the subdwarfs lie within, or just slightly outside, the band defined by
the isochrones.

In fact, there is more than just qualitative consistency as the temperatures
predicted by the evolutionary computations agree rather well with those derived
by CRMBA.  The top panel of Fig.~\ref{fig:sbdte} plots, for each subdwarf, the
difference between the IRFM $\teff$ and the temperature predicted by an
isochrone that has the same metallicity as the star (assuming the [Fe/H] value
given by CRMBA), at its observed $M_V$\ value.  (The $\teff$\ predicted by the
models is readily obtained by interpolating within the grid of isochrones.)  As
indicated, the mean offset between the two is only 42~K, in the sense that the
empirical temperatures tend to be cooler than those predicted by the models,
with a standard deviation of 52~K.  For 11 of the 14 subdwarfs, the {\it observed
minus isochrone} (`Obs--Iso') $\teff$\ differences are less than 50~K.

Alternatively, one can determine the difference between the subdwarf [Fe/H]
value (from CRMBA) and that of the isochrone on which the subdwarf is located.
As shown in the middle panel, the mean `Obs--Iso' difference amounts to $-0.19$
dex, with a standard deviation of 0.25 dex, though the two estimates agree to
within $\sim\pm 0.15$ dex for the majority of the stars in the sample.  However,
it should be appreciated that a small shift in the adopted temperature of any of
the subdwarfs could imply a fairly large offset in its inferred [Fe/H] value
from stellar models because the horizontal ($\log\teff$) separations between
isochrones for metallicities that differ by 0.2 dex are small, especially at 
the lowest [Fe/H] values.  At [Fe/H] $< -2$, a small change in $\teff$ implies
a large difference in the inferred metallicity.  Our point here is that the
{\it scale} of the $\delta$\,[Fe/H] variations may not be particularly
meaningful.  Obviously, the results presented in the top and middle panels are
highly correlated.

In order to enlarge the sample of Population II stars in the solar neighborhood
that can be used as standard candles, \citet{cmo17} determined very accurate
and precise parallaxes for eight nearby subdwarfs with [Fe/H] $\lta -1.4$ using
the {\it HST} Fine Guidance Sensors.  The temperatures and metal abundances of
these stars were determined by \citet{ogc17} via a differential chemical
abundance analysis (relative to the Sun) of high-resolution, high signal-to-noise
spectra.  However, they found that the adoption of the color--$\teff$\ relations
given by CRMBA resulted in steep slopes of the Fe$\,$I abundances with excitation
potential, and that, to remove such trends, it was necessary to adopt cooler
temperatures by as much as $\sim 400$~K than the photometric $\teff$
determinations.  As shown in Fig.~\ref{fig:sbdte}, where six of the subdwarfs
are plotted as open circles\footnote{The other two stars are much brighter, and
as one of them seems to be a subgiant while the other has a questionable
reddening (according to Chaboyer et al.), they have not been plotted.}, such low
temperatures are highly problematic.  In fact, the binaries in M\,55 and
NGC\,6362 (recall Figs.~\ref{fig:mrhr} and \ref{fig:bin4}) completely rule them
out.

Compelling additional support for this conclusion is provided by the fact that
the low $\teff$\ values derived by \citet{ogc17} for 4 of the stars in their
sample place them in close proximity to a 4.3 Gyr, solar abundance isochrone
that matches the MS of M\,67 on the $[(V-K)_S,\,M_V]$-plane down to $M_V \sim
8.5$; see \citet[their Fig.~9]{vbf14}.  Indeed, the part of the isochrone that
has been plotted is indistinguishable from one for 4.55 Gyr (the solar age)
that passes through the location of the Sun on the same CMD or on the
$(\log\teff,\,M_V)$-diagram.  It is simply not possible that field subdwarfs
with [Fe/H] $\lta -1.8$ have the same temperatures as MS stars in M\,67 at the
same absolute magnitudes.  According to the constraints provided by the cluster
binaries considered in this paper, the subdwarfs analyzed by O'Malley et
al.~cannot be much cooler than the predictions of stellar models for their
metallicities (or IRFM-based temperatures), though they could be hotter than
such estimates.

Apparently, the $\teff$\ derived by \citet{ctb12} for HD\,103095 ([Fe/H] $\sim
-1.4$) from CHARA interferometric data suffers from the same problem (note the
location of the filled square in Fig.~\ref{fig:sbdte}).  Indeed, concerns with
the CHARA result have been expressed previously by \citet{szm15}, who have found
that such a low $\teff$\ ($4818 \pm 54$~K) is incompatible with all of the
spectroscopic temperature indicators that they have checked (such as H$_\alpha$
line profiles).  As noted by \citet{hjg15}, the CHARA result is especially
perplexing because most spectroscopic and photometric temperatures for
HD\,103095 have generally been in good agreement, favoring a value close to
5100~K. 

The subgiant HD\,140283 ([Fe/H] $\sim -2.5$) is another example of a
Population II star in which the photometric $\teff$\ value is considerably
higher than the fundamental determination (see \citealt{hjg15}, \citealt{vbn14}).
\citet{ctb15} concluded that it is necessary to adopt a much smaller value of
the mixing-length parameter than the solar value in order to explain the low
$\teff$ that they derived from CHARA observations, but the CMDs of globular
clusters with similar metallicities completely rule out this hypothesis (see
\S\,3.2.2 in Paper I).  More importantly, their temperature for HD\,140283 would
be in conflict with the unequivocal result from the eclipsing binary in M\,55 in
support of the IRFM and stellar model $\teff$\ scales.  Thus, there is
substantial evidence (also see \citealt{cpg14}) that some interferometric
results suffer from systematic errors of one kind or another.\footnote{This
suspicion appears to have been confirmed by new interferometry of HD\,140283 and
HD\,103095 by \citet{kwn18}, who derived $\teff = 5787 \pm 48$~K and $5140 \pm
49$~K, in turn, for these two stars.  As reported by \citet{vbn14}, the IRFM 
temperature of HD\,140283 is 5797~K, when the current best estimate of its
reddening, $E(B-V) = 0.004$, is taken into account.  In the case of HD\,103095,
\citet{csa11} obtained $\teff = 5168$~K from calibrations of several
color--$\teff$\ relations.  For both stars, the photometric temperatures differ
by $< 1$\% from the latest interferometric determinations.}

In any case, the adoption of what has turned out to be the wrong temperatures
clearly calls into question the chemical abundances derived by \citet{ogc17},
as well as the determination of GC distances and ages carried out by
\citet{cmo17} on the basis of the subdwarfs in that sample.  (Presumably,
errors in the temperature structures of the model atmospheres used by O'Malley
et al.~are responsible for the dependence of Fe\,I abundance on excitation
potential that led them to favor cool temperatures.)  Although their abundance
analysis should be repeated in order to derive improved metallicities, we note
that several of the same subdwarfs were included in the spectroscopic survey
carried out by \citet{ica12}, who adopted IRFM temperatures.  Their [Fe/H]
determinations are higher than those obtained by O'Malley et al.~by up to $\sim
0.4$ dex, though there could well be other variations in the respective analysis
methods that would serve to increase or decrease such differences. 

\begin{figure}[t]
\plotone{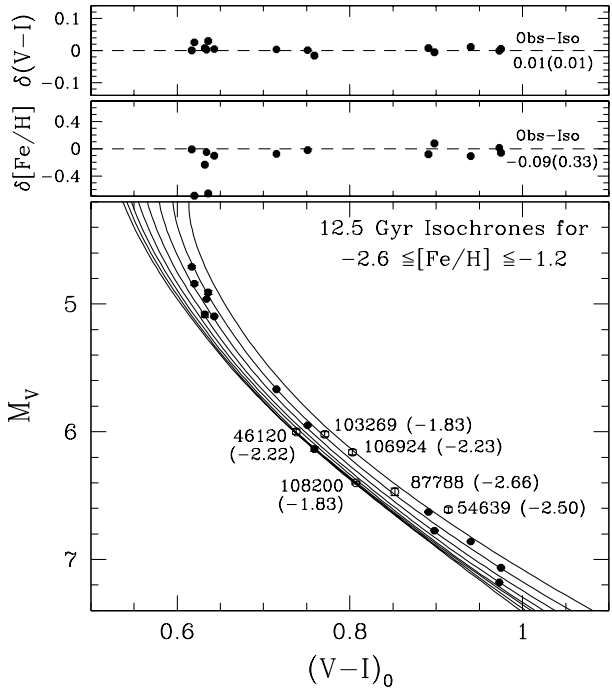}
\caption{Similar to the previous figure, except that the subdwarfs are
compared with isochrones that have been transposed to the $V-I_C$ color plane
using the MARCS color--$\teff$ relations (\citealt{cv14}).  The colors and
metallicities of the subdwarfs have been taken from the study by CRMBA, who
claim that all of these stars are unreddened.}
\label{fig:sbdvi}
\end{figure}

Since the MS-fitting method of determining GC distances involves comparisons of
the apparent magnitudes of cluster stars with the $M_V$\ values of local
subdwarfs at the same intrinsic colors, rather than at common temperatures, it
is instructive to compare the predicted and observed subdwarf properties on
various CMDs.  However, \citet{vcs10} have already shown that MARCS
transformations to $B-V$ are considerably less successful than those for redder
colors in satisfying observational constraints.  Furthermore, redder colors have
the advantage of being less sensitive to metal abundances than $B-V$, and since
[Fe/H] determinations involve substantial uncertainties (recall
Table~\ref{tab:t2} in \S\,\ref{sec:intro}), we decided to focus on the
$[(V-I_C)_0,\,M_V]$-diagram.  It turns out that, as shown in
Figure~\ref{fig:sbdvi}, Victoria-Regina isochrones (\citealt{vbf14}) do a 
particularly fine job of reproducing the $V-I_C$ colors of the same 11 subdwarfs
that were considered in the previous figure when they are transposed from the
theoretical to the observed plane using MARCS color--$\teff$\ relations (from
\citealt{cv14}); in both plots, they are represented by filled circles.  The
mean color offset turns out to be 0.01 mag and the standard deviation is 0.01
mag (see the upper panel).  The differences between the [Fe/H] values given by
CRMBA and those inferred from the isochrones are also small (see the middle
panel).

Because the $\alpha$-element abundances affect both the predicted temperatures
of stars (e.g., see \citealt{vbf14}) and stellar color--$\teff$\ relations
(\citealt{cv14}), the models that are compared with observations should assume
the correct values of [$\alpha$/Fe] (though this is less of a concern at low
metallicities and for $V-I_C$, rather than $B-V$, colors).  As far as we have
been able to determine, the subdwarfs used in this study have [$\alpha$/Fe]
values quite close to $+0.4$, and we have therefore made use of isochrones for
[$\alpha$/Fe] $= +0.4$ in Figs.~\ref{fig:sbdte} and \ref{fig:sbdvi}.  For
instance, \citet{cpf06} give [$\alpha$/Fe] values for HD\,25329, HD\,31128,
HD\,34328, HD\,94028, and HD\,145417 that vary from 0.33 to 0.49, with a mean
value of 0.40, while the analysis of a high-resolution spectrum for HD\,19445
by P.~E.~Nissen (see \citealt{vbn14}) yielded [$\alpha$/Fe] $= 0.39$.

The subdwarfs from the study by \citet{cmo17}, which are plotted in
Fig.~\ref{fig:sbdvi} as open circles and identified by their {\it Hipparcos}
catalog (HIP) numbers, do not line up in the way that they probably should.
Even though the [Fe/H] values given by Chaboyer et al.~(the numbers enclosed by
parentheses) are almost certainly too low, it should still be the case that
HIP\,54639 and HIP\,87788 are the most metal-deficient stars in their sample,
and yet their locations on the $[(V-I_C)_0,\,M_V]$-diagram imply that they have
significantly higher metallicities than HIP\,46120 and HIP\,108200.  (Differences
in the $\alpha$-element abundances could partially explain this, but Chaboyer
et al.~report that HIP\,54639 has a lower value of [$\alpha$/Fe] than
HIP\,46120, which is in the wrong sense to explain the offset between these
two stars.)  Also, HIP\,103269 and HIP\,108200 should lie on the same isochrone
if they have nearly the same [Fe/H] and [$\alpha$/Fe] values, as reported by
Chaboyer et al.~(something that needs to be checked by a follow-up spectroscopic
study), but they do not do so.  Even though Chaboyer have obtained very accurate
distances for the subdwarfs in their sample, they are clearly of limited
usefulness as standard candles at the present time because of problems with
their effective temperatures and metallicities.

Interestingly, the most discrepant points in Fig.~\ref{fig:sbdte} (e.g., the
one representing HD\,25329) are much less problematic in Fig.~\ref{fig:sbdvi}.
Their $(V-I)_C$ colors suggest, in fact, that the temperatures given by CRMBA
for these stars may be too cool.  Anyway, increasing their $\teff$\ values by
$\sim 100$~K, which is within the the $2\,\sigma$\ uncertainties of the
temperatures derived by CRMBA, would make their locations in the
$[(\log\teff,\,M_V]$-diagram much more consistent with the properties of the
other subdwarfs in our sample. 

\subsubsection{The Subdwarf-Based Distance Modulus of M\,55}
\label{subsubsec:sbdism55}

Fortunately, \citet{mfr96} have obtained deep $VI_C$ photometry for M\,55, and
we can therefore fit the MS fiducial that they derived to the local subdwarfs
in order to determine the cluster distance modulus.  We checked that their 
photometry is in good agreement with the observations that are publicly
available for M\,55 in P.~Stetson's ``Photometric Standard Fields" archive (see
footnote 3), and we found that the level of consistency between the two data
sets is quite satisfactory.  We then fitted a cubic to the fiducial points given
by Mandushev et al.~(from their Table 3), over the range in $V$ from 18.8 to
21.4, using a standard least-squares fitting code.  When matching the cluster MS
to the subdwarfs in the solar neighborhood, we will assume that the subdwarfs
define a cubic with exactly the same linear, quadratic, and cubic coefficients
as the one that has been fitted to the cluster photometry.  The difference in
the zero points (the constant terms) is $V-M_V$, which is the apparent distance
modulus that we are seeking.

To produce a mono-metallicity subdwarf sequence for [Fe/H] $= -1.85$, which
represents our best estimate of the metallicity of M\,55 (see
Fig.~\ref{fig:m55fit}), the observed $V-I_C$ color for each of the 11 subdwarfs
is adjusted by the difference in color, at its absolute $V$ magnitude, between
an isochrone for [Fe/H] $= -1.85$ and one for the metallicity of the star.
Thus, the colors of subdwarfs that are somewhat more metal-rich than M\,55 will
be shifted to slightly bluer $V-I_C$ colors, and vice versa.  Note that the
models are used in only a differential sense to determine these adjustments,
which range from $-0.026$\ mag in the case of HD\,103095 and HD\,145417 (the
most metal-rich stars in our sample) to $+0.012$\ mag in the case of
BD$+$023375, which has the lowest metallicity.  These corrections are quite
minor because the horizontal separations between isochrones for different [Fe/H]
values are small, which is the main (and important) advantage of performing this
analysis on the $[(V-I_C)_0,\,M_V]$-diagram instead of the $[(B-V)_0,\,M_V]$-plane.  

\begin{figure}[t]
\plotone{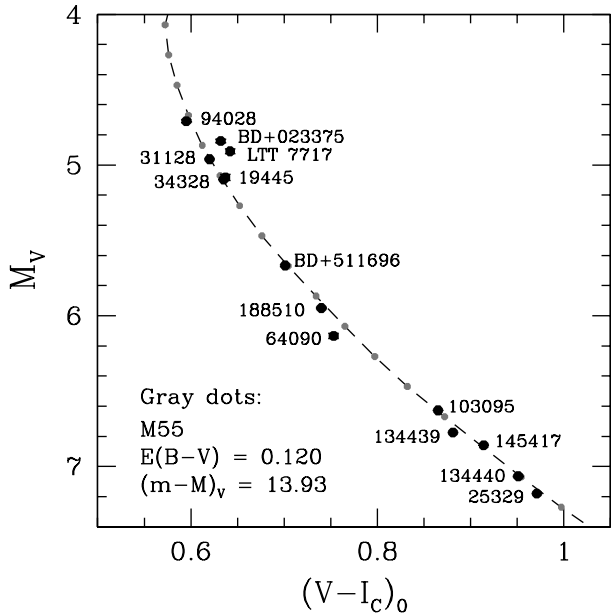}
\caption{Fit of the main-sequence fiducial of M\,55 (gray dots connected by a
dashed curve) to the 11 subdwarfs in our sample (filled circles with vertical
error bars), once small corrections have been applied to their colors to 
compensate for the difference between their metallicities and that of the
cluster (which is assumed to have [Fe/H] $= -1.85$).  If $E(B-V) = 0.120$, the
MS fit yields $(m-M)_V = 13.93$ (as indicated).}
\label{fig:m55sbd}
\end{figure}

Once the mono-metallicity subdwarf sequence has been defined, one can use, e.g.,
the LFIT subroutine in {\it Numerical Recipes} (\citealt{ptv07}) to determine
the zero-point of the cubic equation that provides the optimum fit to the
subdwarfs, assuming that it has exactly the same {\it shape} as the one that has
been fitted to the cluster MS.  This particular computer program uses $\chi^2$
minimization to determine the value of the constant term, with each star given
a weighting of $1/\sigma(M_V)^2$.  It turns out that $V-M_V = 13.93$ if M\,55
has [Fe/H] $= -1.85$ and $E(B-V) = 0.12$; see Figure~\ref{fig:m55sbd}, which
illustrates that the subdwarfs follow the morphology of the cluster MS fiducial
rather well.

This determination depends much more on the reddening of M\,55 than its
metallicity.  For instance, we would obtain $(m-M)_V = 14.00$ if $E(B-V) = 0.13$
and [Fe/H] $= -1.85$, as compared with 13.91 if $E(B-V) = 0.12$ and [Fe/H] $=
-2.00$.  All of these results agree rather well with those found from fits of
ZAHB models to the cluster HB stars (see Figs.~\ref{fig:hbvas} and
\ref{fig:bvm18}), CMD simulations of the observed HB population
(Fig.~\ref{fig:syn6809}), and comparisons of the predicted and measured periods
of the cluster RR Lyrae variables (see Table~\ref{tab:t3}).  Thus, nearly the
same distance modulus is obtained for M\,55 whether it is based on nearby
subdwarfs, or the evolutionary and pulsational properties of the cluster HB
stars.  (As deep, high-quality $VI_C$ photometry has not yet been
obtained for NGC\,6362, as far as we are aware, we are not able to carry out
a similar determination of its distance modulus as that just described for
M\,55.)

\section{Summary}
\label{sec:sum}

This investigation has presented a detailed analysis of the evolutionary and
pulsational properties of the stars in M\,55 and NGC\,6362.  These GCs were
selected for the subject of this study because they contain detached, totally
eclipsing binary stars with well determined masses and radii that can be used
to constrain the chemical properties of the two clusters. Even more importantly,
binaries are able to set tight limits on the $\teff$\ scale of Population II
stars, if their distances (and hence their luminosities) can be established to
high accuracy and the uncertainties of their measured radii are small.  

In this series of papers, we have used ZAHB models, simulated HBs, and
comparisons of the predicted and observed periods of RR Lyrae variables to set
the cluster distances.  It is well known that the luminosities of HB stars
depend on the envelope helium abundance, $Y_{\rm env}$, and the mass of the
helium core, ${\cal M}_c^{\rm He}$.  Metal-deficient GCs cannot have an initial
He abundance, $Y_0$, that is less than the primordial value of $Y$ and, if
diffusive processes are treated, the abundance of He in the convective envelopes
of upper-RGB stars (i.e., after the first dredge-up) will be the lowest
abundance that is predicted by stellar models ($Y_{\rm env} \lta Y_0$).  Because
RGBT models that neglect diffusion have values of $Y_{\rm env}$\ that are larger
than $Y_0$\ by $\sim 0.01$--0.02 (see, e.g., \citealt[their Table 2]{swc17}), the
corresponding HB models will necessarily be brighter than those that treat this
physics.  (The fact that essentially the same distance modulus is obtained for
M\,3 from our HB models and from the best available calibration of the RR Lyrae
standard candle provides a strong argument that diffusive processes cannot be
ignored.)  Moreover, any extra mixing processes that enhance $Y_{\rm env}$, or
any non-canonical effects (e.g., rotation) that increase ${\cal M}_c^{\rm He}$,
will also result in brighter HBs.

Based on these considerations, we would argue that the cluster distance moduli
that we have derived from our HB computations cannot be much smaller than our
determinations. (In fact, our estimates of $(m-M)_V$ appear to be quite accurate
given that, in particular, a fit of the MS of M\,55 to solar-neighborhood
subdwarfs yields the same distance modulus for this GC to within $\sim 0.03$
mag.)  On the assumption of the relatively short distance moduli obtained in
this study, the temperatures that we have derived (from ${\cal L} =
4\pi R^2\sigma\teff^4)$ for the binaries in M\,55 and NGC\,6362 agree very well
with those predicted by our isochrones --- and with those inferred from the
calibration of the IRFM by CRMBA.  Any increase in the adopted values of
$(m-M)_V$\ would result in hotter temperatures.  

This finding rules out the very cool temperatures that were found in the recent
spectroscopic study of several field subdwarfs by \citet{ogc17}, for instance,
as well as those derived in some interferometric studies (e.g.,
\citealt[2015]{ctb12}).  Such cool $\teff$\ values as those reported in these
investigations for stars with [Fe/H] $\lta -1.3$ present the additional
difficulty that their location on the H-R diagram would place some of them on,
or adjacent to, a solar-abundance, solar-age isochrone that satisfies the solar
constraint.  It is clear that determinations of the chemical abundances of
Pop.~II stars from their spectra {\it must} adopt relatively warm temperatures
close to those favored by stellar models and the IRFM.  Because detached,
totally eclipsing binaries in GCs with well determined distances provide such
powerful constraints on the $\teff$\ scales, mass-radius relations, and
mass-luminosity relations that apply to those clusters, efforts to discover
and to observe such binaries should be given high priority and strong support.
Once the temperatures are known, improvements to, among other things, metal
abundance determinations, color--$\teff$\ relations, the temperature structures
of model atmospheres, and (perhaps) to our understanding of convection theory,
diffusion, and any other physics that can have a significant impact on the
temperatures of stellar models will follow.

Our HB simulations indicate that both M\,55 and NGC\,6362 contain stars that
span a small range in $Y$.  We are able to reproduce the morphology and
distribution of stars along the HB in M\,55 if approximately 
half of them have $Y \approx 0.25$ and only $\sim 10$\% have $Y \gta 0.27$.
The CMD for the HB stars in NGC\,6362 is unusual insofar as the reddest
non-variable stars are significantly fainter than the faintest RR Lyrae and
non-variable stars just to the blue of the instability strip.  These
observations are readily explained if there are similar numbers of stars with
$Y = 0.25$, 0.265, and 0.28, and all of the RR Lyrae and bluer HB stars have
enhanced helium abundances.  Because the populations of HB stars with different
$Y$ span such a wide color range, the luminosity offsets between them are
easily seen, in contrast with the HBs of both more metal-rich and more
metal-poor GCs, such as 47 Tuc (see Paper II) and M\,55 (this paper).  As a
result, NGC\,6362 provides one of the most striking examples of the existence
of multiple stellar populations in globular clusters.  Encouragingly, the
reddest $ab$-type pulsators in NGC\,6362 appear to follow the morphology
(notably the blue loops) of post-ZAHB evolutionary tracks quite well.

Due to the spread in $Y$, the mass-radius and mass-luminosity diagrams for
the binary components do not constrain the cluster metallicities particularly
well because the effects of $\delta Y = 0.02$ on predicted ${\cal M}$--${\cal R}$
and ${\cal M}$--$M_V$ relations is larger than the effects of a $+0.2$\ dex
change in [Fe/H].  As a result, the metallicity that is inferred from such plots
depends on whether a particular binary belongs to the helium-normal or
helium-enhanced population.  Nevertheless, the binaries indicate that M\,55 and
NGC\,6362 have [Fe/H] $\approx -1.85$\ and $\approx -0.9$, respectively, with
$1\sigma$\ uncertainties amounting to $\sim 0.1$\ dex.  On the other hand, since
we are using the HB to set the cluster distances, and the difference in the
luminosity of the HB and the turnoff, \delv, to determine the cluster age,
the uncertainty in the metallicity is not a serious concern; at higher [Fe/H]
values, both the HB and the turnoff at a fixed age become fainter (though not
by exactly the same amounts), and vice versa.  Hence, as long as there is a
significant population of stars with $Y = 0.25$, to justify the use of HB models
for this He abundance in determining the cluster distance, the effect of a 0.2
dex uncertainty in the metallicity on the derived age will be relatively minor.

If we adopt [O/Fe] $= 0.5 \pm 0.1$ and [$m$/H] $= 0.4$\ for the other
$\alpha$-elements, we obtain $(m-M)_V = 13.95 \pm 0.05$ and an age of $12.9
\pm 0.8$\ Gyr for M\,55, as compared with $(m-M)_V = 14.56 \pm 0.05$ and $12.4
\pm 0.8$\ Gyr for NGC\,6362.  The distance modulus uncertainty, which
contributes $\approx \pm 0.5$\ Gyr to the age uncertainty, corresponds to the
range of possible values for which our models are able to reproduce the mean
periods of the $ab$- and $c$-type RR Lyrae to within $\sim \pm 0.03$~days.  The 
effects of chemical abundance uncertainties account for the remainder of the age
uncertainty.  The ages derived here for M\,55 and NGC\,6362 should be
particularly robust because they are based on well-tested, up-to-date stellar
models and they satisfy the constraints provided by member eclipsing binaries
and RR Lyrae variables, as well as solar neighborhood subdwarfs (which were
considered only in the case of M\,55).

\acknowledgements

We are particularly grateful to M\'arcio Catelan for the many comments and
suggestions that he provided as a result of his very careful reading of the
first draft of the manuscript.  This paper also benefitted from discussions
with Karsten Brogaard and from a thoughtful and helpful report by an
anonymous referee.  Financial support for this investigation was provided by
a Discovery Grant to D.A.V.~from the Natural Sciences and Engineering Research
Council of Canada.  This work has made use of data from the European Space
Agency (ESA) mission {\it Gaia} (\url{https://www.cosmos.esa.int/gaia}),
processed by the {\it Gaia} Data Processing and Analysis Consortium (DPAC, 
\url{https://www.cosmos.esa.int/web/gaia/dpac/consortium}).  Funding for the
DPAC has been provided by national institutions, in particular the 
institutions participating in the {\it Gaia} Multilateral Agreement.

\end{document}